\documentclass[a4paper,11pt]{article}
\pdfoutput=1 
\usepackage{jheppub} 
\usepackage[T1]{fontenc} 
\usepackage[utf8]{inputenc}
\usepackage{amsfonts}
\usepackage{graphics}
\usepackage{epsfig} 
\usepackage[english]{babel} 
\usepackage{caption} 
\usepackage{subcaption} 
\usepackage{amsmath}
\usepackage{bm}
\usepackage{multirow}
\usepackage{array,booktabs}
\usepackage{calc}
\usepackage[table]{xcolor}
\usepackage{color}  
\usepackage{makecell} 

\definecolor{aqua}{rgb}{0.0, 1.0, 1.0}
\definecolor{babyblue}{rgb}{0.54, 0.81, 0.94}
\definecolor{beaublue}{rgb}{0.74, 0.83, 0.9}
\definecolor{blizzardblue}{rgb}{0.93, 0.93, 0.93} 
\definecolor{cyan}{rgb}{0.0, 1.0, 1.0}
\newcolumntype{?}{!{\vrule width 0.8pt}}

\def\sinb{\sin\beta}
\def\cosb{\cos\beta}
\def\sina{\sin\alpha}
\def\cosa{\cos\alpha}
\def\tanb{\tan\beta}
\def\hpm{H^\pm}
\def\mhpm{m_{\hpm}}
\def\mh{m_h}
\def\mH{m_H}
\def\ma{m_A}
\def\tb{\tan\beta}
\def\gaga{\gamma\gamma}
\def\cb{\cot\beta}
\def\bma{\beta-\alpha}  

\newcommand{\RN}[1]{
  \textup{\uppercase\expandafter{\romannumeral#1}}%
\renewcommand\thesubfigure{(\alph{subfigure})} 
\captionsetup[sub]{labelformat=simple} 
}
 
\title{\boldmath Capability of future linear colliders to discover heavy neutral CP-even and CP-odd Higgs bosons within Type-\RN{1} 2HDM} 
\author[]{Majid Hashemi and Gholamhossein Haghighat}
\affiliation[]{Physics Department, College of Sciences, Shiraz University, \\ Shiraz, 71946-84795, Iran}
\emailAdd{hashemi$\_$mj$@$shirazu.ac.ir, hosseinhaqiqat$@$gmail.com} 

\abstract{In this study, assuming the type-\RN{1} 2HDM at SM-like scenario, observability of heavy neutral CP-even and CP-odd Higgs bosons $H$ and $A$ at a linear collider operating at $\sqrt{s}=1$ TeV is investigated through the signal process chain $e^- e^+ \rightarrow A H \rightarrow ZHH$ where the $Z$ boson experiences a leptonic ($e^-e^+$ or $\mu^-\mu^+$) decay and the $H$ Higgs boson is assumed to decay into a di-photon. This signal process is motivated especially by the clean signature which leptonic and photonic events can provide at colliders, and also by the enhancement due to the charged Higgs-mediated contribution to $H$ di-photon decay at large $\tb$ values. Simulation is based on four benchmark points according to which the Higgs mass $m_H$ ($m_A$) varies within the range $150$-$300$ ($200$-$400$) GeV. Energy smearing of jets and photons are performed, and momentum smearing is also applied to leptons. Results indicate that, for all of the assumed benchmark points, the Higgs bosons $H$ and $A$ are observable with measurable masses, and with signals exceeding $5\sigma$ at integrated luminosities $111$ and $201\,fb^{-1}$ respectively. Such luminosities are easily accessible to future linear colliders. 
 }      

\begin{document}
\maketitle 
\flushbottom
 
\section{Introduction}
The success of the standard model of elementary particles in explaining a wide range of phenomena and the experimental verification of the Higgs boson \cite{HiggsObservationCMS,HiggsObservationATLAS} which had been theorised \cite{Englert1,Higgs1,Higgs2,Kibble1,Higgs3,Kibble2} many years before its discovery have been important motivations behind the idea of extending the standard model, an idea which may pave the way for new physics to resolve the present challenging issues of science. During the last several decades, various kinds of extensions of the standard model have emerged as significant candidates for new physics. 
 
The standard model (SM) employs the simplest possible scalar structure and consequently predicts one Higgs boson resulting from the single assumed $SU(2)$ Higgs doublet. However, axion models \cite{KIM1}, supersymmetry \cite{MSSM1}, the SM inability to explain the baryon asymmetry of the universe \cite{Trodden}, etc., have motivated people to add another $SU(2)$ doublet to the SM scalar structure. Two-Higgs-doublet model (2HDM) \cite{2hdm_TheoryPheno,2hdm1,2hdm2,2hdm3,2hdm4_CompositeHiggs,2hdm_HiggsSector1,2hdm_HiggsSector2,Campos:2017dgc}, as one of the simplest extensions of the SM, makes use of two $SU(2)$ Higgs doublets. Employing two doublets in this model, immediately leads to the prediction of five Higgs bosons, one of which (the lightest one) is assumed to be the same as the observed SM Higgs boson, and the others are assumed to be undiscovered Higgs bosons which may be observed in future. Two out of four undiscovered Higgs bosons are neutral CP-even and CP-odd Higgs bosons $H$ and $A$, and the other two are charged Higgs bosons $H^\pm$. In this study, observability of the neutral Higgs bosons $H$ and $A$ at a linear collider operating at $\sqrt{s}=1$ TeV is addressed. 

Imposing the discrete $Z_2$ symmetry results in four types for 2HDM which naturally conserve flavor. In this work, the type \RN{1} at SM-like scenario is assumed as the theoretical framework and the process $e^- e^+ \rightarrow A H\rightarrow ZHH$ followed by the decays $H\rightarrow \gamma\gamma$ and $Z\rightarrow e^-e^+$ or $\mu^-\mu^+$ is chosen as the signal process. Such a signal process serves experimentalists well in search for Higgs bosons since the final state photons and leptons provide a simple and clear signature at linear colliders. Furthermore, the large enhancement due to the charged Higgs-mediated contribution to the $H$ di-photon decay at large $\tb$ values considerably boosts the signal cross section and has been an important motivation behind this signal process.
  
In comparison with the SM and Minimal Supersymmetric Standard Model (MSSM) \cite{MSSM1,MSSM2,MSSM3} which constrains the Higgs masses, exploring whole parameter space of the 2HDM takes much longer because of its larger number of free parameters. In this work, assuming four benchmark points in the parameter space of the type-\RN{1} 2HDM, observability of the heavy neutral scalar and pseudoscalar Higgs bosons $H$ and $A$ is studied. Applying appropriate selection cuts, reconstructed masses of the Higgs bosons will be obtained with few GeV uncertainty and it will be shown that, for all of the assumed benchmark points, the Higgs bosons $H$ and $A$ are observable with signals exceeding $5\sigma$ at integrated luminosities $111$ and $201\,fb^{-1}$ respectively.  

\section{Two-Higgs-doublet model}
Extending the standard model scalar structure by introducing another $SU(2)$ Higgs doublet and employing the general Higgs potential 
\begin{equation}
  \begin{aligned}
    \mathcal{V} = &m_{11}^2\Phi_1^\dagger\Phi_1+m_{22}^2\Phi_2^\dagger\Phi_2
    -\left[m_{12}^2\Phi_1^\dagger\Phi_2+\mathrm{h.c.}\right]
    \\
    &+\frac{1}{2}\lambda_1\left(\Phi_1^\dagger\Phi_1\right)^2
    +\frac{1}{2}\lambda_2\left(\Phi_2^\dagger\Phi_2\right)^2
    +\lambda_3\left(\Phi_1^\dagger\Phi_1\right)\left(\Phi_2^\dagger\Phi_2\right)
    +\lambda_4\left(\Phi_1^\dagger\Phi_2\right)\left(\Phi_2^\dagger\Phi_1\right)
    \\&+\left\{
    \frac{1}{2}\lambda_5\left(\Phi_1^\dagger\Phi_2\right)^2
    +\left[\lambda_6\left(\Phi_1^\dagger\Phi_1\right)
      +\lambda_7\left(\Phi_2^\dagger\Phi_2\right)
      \right]\left(\Phi_1^\dagger\Phi_2\right)
    +\mathrm{h.c.}\right\},
  \end{aligned}
  \label{lag}
\end{equation}
where 
\begin{equation}
\begin{gathered}
\Phi_i=\left(  
\begin{array}{c}
\phi_i^+ \cr  ( v_i+\rho_i +i \eta_i)/\sqrt 2 
\end{array}
\right)\,\quad  i=1,2 \\
v_1=v\cosb, \quad v_2=v \sin \beta, \quad v=(\sqrt{2}G_F)^{-1/2}\approx246\ GeV ,
\end{gathered}
\end{equation}
the general two-Higgs-doublet model forms and leads to the prediction of three neutral Higgs bosons $h$, $H$ and $A$, and two charged Higgs bosons $H^\pm$. The physical scalars 
\begin{equation}
h=-\rho_1\sina + \rho_2\cosa,\quad H=\rho_1\cosa + \rho_2\sina ,
\end{equation}
represent the neutral CP-even Higgs bosons $h$ and $H$, and the physical pseudoscalar 
\begin{equation}
A=-\eta_1\sinb+\eta_2\cosb, 
\end{equation}
represents the neutral CP-odd Higgs boson $A$. By convention and without loss of generality $0\leq \beta \leq \pi/2$ and  $-\pi/2\leq \alpha \leq \pi/2$ are chosen. Working in the so-called ``physical basis'', the physical Higgs masses ($\mH,\mh,\ma,\mhpm$), the Higgs v.e.v.'s ratio ($\tanb$), the CP-even Higgs mixing angle ($\alpha$), $m_{12}^2$, $\lambda_6$ and $\lambda_7$ must be determined to fully specify the Higgs potential \cite{2hdm_TheoryPheno}. The values of $m_{11}^2$ and $m_{22}^2$ are determined by the minimization conditions for a minimum of the vacuum once $\tan\beta$ is determined. To avoid tree level flavour-changing neutral currents (FCNC), the discrete $Z_2$ symmetry ($\Phi_1\to \Phi_1$ and $\Phi_2\to -\Phi_2$) is imposed \cite{2hdm2,2hdm3,2hdm4_CompositeHiggs}, and as a result, the parameters $\lambda_6$ and $\lambda_7$ are set to zero. Allowing a non-zero value for the parameter $m_{12}^2$, the $Z_2$ symmetry is softly broken though. In general, the parameters $m_{12}^2$ and $\lambda_5$ are complex. However, assuming CP invariance, they are taken to be real in this paper.

As a result of the imposed $Z_2$ symmetry, Higgs coupling to fermions is constrained to follow the patterns provided in table \ref{coupling}.
\begin{table}[h]
\normalsize
\fontsize{11}{7.2} 
    \begin{center}
         \begin{tabular}{ >{\centering\arraybackslash}m{.6in} >{\centering\arraybackslash}m{.4in} >{\centering\arraybackslash}m{.4in} >{\centering\arraybackslash}m{.4in}   }
& \cellcolor{blizzardblue}{$u_R^i$} & \cellcolor{blizzardblue}{$d_R^i$} & \cellcolor{blizzardblue}{$\ell_R^i$} \parbox{0pt}{\rule{0pt}{1ex+\baselineskip}}\\ \Xhline{3\arrayrulewidth}
 \cellcolor{blizzardblue}{Type \RN{1}} &$\Phi_2$ &$\Phi_2$ &$\Phi_2$ \parbox{0pt}{\rule{0pt}{1ex+\baselineskip}}\\ 
 \cellcolor{blizzardblue}{Type $\RN{2}$} &$\Phi_2$ &$\Phi_1$ &$\Phi_1$ \parbox{0pt}{\rule{0pt}{1ex+\baselineskip}}\\ 
\cellcolor{blizzardblue}{Type X} &$\Phi_2$ &$\Phi_2$ &$\Phi_1$  \parbox{0pt}{\rule{0pt}{1ex+\baselineskip}}\\ 
\cellcolor{blizzardblue}{Type Y} &$\Phi_2$ &$\Phi_1$ &$\Phi_2$  \parbox{0pt}{\rule{0pt}{1ex+\baselineskip}}\\ \Xhline{3\arrayrulewidth}
  \end{tabular}
\caption{Higgs coupling to up-type quarks, down-type quarks and leptons in types with natural flavour conservation. The superscript $i$ is a generation index.  \label{coupling}}
  \end{center}
\end{table}
Accordingly, there are four types of 2HDM which naturally conserve flavour. The types ``X'' and ``Y'' are also called ``lepton-specific'' and ``flipped'' respectively. Following table \ref{coupling}, Higgs-fermion interaction Lagrangian of the 2HDM takes the form \cite{2hdm_TheoryPheno}
\begin{equation}
\begin{aligned}
\mathcal{L}_{\ Yukawa}\ =\ & -\ \sum_{f=u,d,\ell}\ \dfrac{m_f}{v}\ \Big(\xi_{h}^{f}\bar{f}fh\ +\ \xi_{H}^{f}\bar{f}fH\ -\ i\xi_{A}^{f}\bar{f}\gamma_5fA \Big)\\
&-\ \Bigg\{\dfrac{\sqrt{2}V_{ud}}{v}\bar{u}\ \big(m_u\xi_A^uP_L\ +\ m_d\xi_A^dP_R\big)\ dH^+\ +\ \dfrac{\sqrt{2}m_\ell\xi_A^\ell}{v}\overline{\nu_L}\ell_RH^+\ +\ H.c. \Bigg\},
\label{yukawa1}
\end{aligned}
\end{equation}
where $P_{L/R}$ are projection operators for left/right-handed fermions and $\xi^X_Y$ factors corresponding to different types are presented in table \ref{xi}.
\begin{table}[h]
\normalsize
\fontsize{11}{7.2} 
    \begin{center}
         \begin{tabular}{ >{\centering\arraybackslash}m{.23in} >{\centering\arraybackslash}m{.55in}  >{\centering\arraybackslash}m{.55in} >{\centering\arraybackslash}m{.55in}  >{\centering\arraybackslash}m{.55in}  }
& \cellcolor{blizzardblue}{\RN{1}} & \cellcolor{blizzardblue}{$\RN{2}$} & \cellcolor{blizzardblue}{X} & \cellcolor{blizzardblue}{Y} \parbox{0pt}{\rule{0pt}{1ex+\baselineskip}}\\ \Xhline{3\arrayrulewidth}
 \cellcolor{blizzardblue}{$\xi_h^u$} &$c_\alpha/s_\beta$ &$c_\alpha/s_\beta$ &$c_\alpha/s_\beta$  &$c_\alpha/s_\beta$  \parbox{0pt}{\rule{0pt}{1ex+\baselineskip}}\\ 
 \cellcolor{blizzardblue}{$\xi_h^d$} &$c_\alpha/s_\beta$ &$-s_\alpha/c_\beta$ &$c_\alpha/s_\beta$  &$-s_\alpha/c_\beta$  \parbox{0pt}{\rule{0pt}{1ex+\baselineskip}}\\ 
 \cellcolor{blizzardblue}{$\xi_h^\ell$} &$c_\alpha/s_\beta$ &$-s_\alpha/c_\beta$ &$-s_\alpha/c_\beta$   &$c_\alpha/s_\beta$  \parbox{0pt}{\rule{0pt}{1ex+\baselineskip}}\\ 
 \cellcolor{blizzardblue}{$\xi_H^u$} &$s_\alpha/s_\beta$ &$s_\alpha/s_\beta$ &$s_\alpha/s_\beta$  &$s_\alpha/s_\beta$  \parbox{0pt}{\rule{0pt}{1ex+\baselineskip}}\\ 
 \cellcolor{blizzardblue}{$\xi_H^d$} &$s_\alpha/s_\beta$ &$c_\alpha/c_\beta$ &$s_\alpha/s_\beta$  &$c_\alpha/c_\beta$  \parbox{0pt}{\rule{0pt}{1ex+\baselineskip}}\\ 
 \cellcolor{blizzardblue}{$\xi_H^\ell$} &$s_\alpha/s_\beta$ &$c_\alpha/c_\beta$ &$c_\alpha/c_\beta$  &$s_\alpha/s_\beta$  \parbox{0pt}{\rule{0pt}{1ex+\baselineskip}}\\ 
 \cellcolor{blizzardblue}{$\xi_A^u$} &$\cot\beta$ &$\cot\beta$ &$\cot\beta$  &$\cot\beta$  \parbox{0pt}{\rule{0pt}{1ex+\baselineskip}}\\ 
 \cellcolor{blizzardblue}{$\xi_A^d$} &$-\cot\beta$ &$\tan\beta$ &$-\cot\beta$  & $\tan\beta$ \parbox{0pt}{\rule{0pt}{1ex+\baselineskip}}\\ 
 \cellcolor{blizzardblue}{$\xi_A^\ell$} &$-\cot\beta$ &$\tan\beta$ &$\tan\beta$  & $-\cot\beta$ \parbox{0pt}{\rule{0pt}{1ex+\baselineskip}}\\ \Xhline{3\arrayrulewidth}
  \end{tabular}
\caption{Factors $\xi^X_Y$ in different types of the 2HDM ($c_x\equiv\cos x$ and $s_x\equiv\sin x$). \label{xi}}
  \end{center}
\end{table}

Choosing the SM-like scenario by the assumption of $\sin(\bma)=1$ \cite{2hdm_TheoryPheno}, the lighter CP-even Higgs boson $h$ is taken as the SM-like Higgs boson. Consequently, the neutral Higgs part of the Yukawa Lagrangian becomes \cite{Barger_2hdmTypes}
\begin{equation}
\begin{aligned}
\mathcal{L}_{\ Yukawa}\ =\ & -v^{-1}  \Big(\ m_d\ \bar{d}d\ +\ m_u\ \bar{u}u\ +\ m_\ell\ \bar{\ell}\ell\ \Big)\ h \\
      & +v^{-1} \Big(\ \rho^dm_d\ \bar{d}d\ +\ \rho^um_u\ \bar{u}u\ +\ \rho^\ell m_\ell\ \bar{\ell}\ell\ \Big)\ H \\
& +iv^{-1}\Big(-\rho^dm_d\ \bar{d}\gamma_5d\ +\ \rho^um_u\ \bar{u}\gamma_5u\ -\ \rho^\ell m_\ell\ \bar{\ell}\gamma_5\ell\ \Big)\ A,
\label{yukawa2}
\end{aligned}
\end{equation}
 where $\rho^X$ factors corresponding to different types are given in table \ref{rho}.
\begin{table}[h]
\normalsize
\fontsize{11}{7.2} 
    \begin{center}
         \begin{tabular}{ >{\centering\arraybackslash}m{.23in}  >{\centering\arraybackslash}m{.55in}  >{\centering\arraybackslash}m{.55in} >{\centering\arraybackslash}m{.55in}  >{\centering\arraybackslash}m{.55in}}
& \cellcolor{blizzardblue}{\RN{1}} 
& \cellcolor{blizzardblue}{$\RN{2}$} & \cellcolor{blizzardblue}{X} & \cellcolor{blizzardblue}{Y} \parbox{0pt}{\rule{0pt}{1ex+\baselineskip}}\\ \Xhline{3\arrayrulewidth}
 \cellcolor{blizzardblue}{$\rho^d$} &$\cot{\beta}$ &$- \tan\beta$ &$\cot\beta$ &$-\tan\beta$ \parbox{0pt}{\rule{0pt}{1ex+\baselineskip}}\\ 
\cellcolor{blizzardblue}{$\rho^u$} &$\cot{\beta}$ &$\cot\beta$ &$\cot\beta$ &$\cot\beta$  \parbox{0pt}{\rule{0pt}{1ex+\baselineskip}}\\   
\cellcolor{blizzardblue}{$\rho^\ell$} &$\cot{\beta}$ &$- \tan\beta$ &$-\tan\beta$&$\cot\beta$  \parbox{0pt}{\rule{0pt}{1ex+\baselineskip}}\\ \Xhline{3\arrayrulewidth} 
 \end{tabular}
\caption{$\rho^X$ factors of the Yukawa Lagrangian in different types. \label{rho}}
  \end{center}    
\end{table}  
As seen in table \ref{rho}, different types of the 2HDM acquire different couplings and therefore, are expected to possess different phenomenological characteristics \cite{2hdm_HiggsSector2}. In the type X, the decay of the neutral Higgs boson $H$ into a di-lepton is enhanced at large $\tb$ values as the corresponding coupling depends on $\tb$ according to table \ref{rho}. In the context of this type, the study \cite{H-2HDMX} takes advantage of the leptonic decay mode enhancement in order to reconstruct the Higgs boson $H$ and measure its mass at a linear collider. In the type \RN{1}, all fermionic decays of the Higgs bosons $H$ and $A$ are suppressed for large $\tb$ values since the corresponding couplings depend on $\cb$. Such a suppression at large $\tb$ values along with an enhancement which will be discussed in great detail in the following section leads to significant phenomenological consequences. 

In order to gain some insight into the behaviour of the cross section of the signal process assumed in this paper, a short summary of the $\alpha$-$\beta$ dependency of the Higgs couplings to fermions as well as weak gauge bosons in the context of the type-\RN{1} 2HDM is provided in table \ref{AHcoupling}. 
\begin{table}[h]
\normalsize
\fontsize{11}{7.2} 
    \begin{center}  
         \begin{tabular}{ >{\centering\arraybackslash}m{1.2in} >{\centering\arraybackslash}m{.5in} >{\centering\arraybackslash}m{.62in} >{\centering\arraybackslash}m{.5in} >{\centering\arraybackslash}m{.5in} >{\centering\arraybackslash}m{.72in} }
& \cellcolor{blizzardblue}{$Af\bar{f}$} & \cellcolor{blizzardblue}{$AZH$} & \cellcolor{blizzardblue}{$AVV$} & \cellcolor{blizzardblue}{$Hf\bar{f}$} & \cellcolor{blizzardblue}{$HVV$} \parbox{0pt}{\rule{0pt}{1ex+\baselineskip}}\\ \Xhline{3\arrayrulewidth}
 \cellcolor{blizzardblue}{$\alpha$-$\beta$ dependency} &$\cot\beta$ &$\sin(\bma)$ &$0$  &$\cot\beta$ &$\cos(\bma)$ \parbox{0pt}{\rule{0pt}{1ex+\baselineskip}}\\ 
\Xhline{3\arrayrulewidth}
  \end{tabular}
\caption{$\alpha$-$\beta$ dependency of the relevant type-I 2HDM couplings. \label{AHcoupling}}
  \end{center}
\end{table}
According to table \ref{AHcoupling}, at the chosen SM-like scenario ($\sin(\bma)=1$), decays corresponding to the $AZH$ coupling acquire their maximum possible widths, and on the other hand, the $H$ field becomes gauge-phobic since the $HVV$ coupling (where $V$ is a weak gauge boson) vanishes. The $AVV$ interaction is also absent independently of the chosen values for $\alpha$ and $\beta$. Such properties along with a boost due to the triple Higgs self-coupling (fully described in the following section) results in a large enhancement which the signal process assumed in this study benefits from. The following section is devoted to the description of the signal and background processes.

\section{Signal and background processes}
In this work, the type-\RN{1} 2HDM is chosen as the theoretical framework and the process chain $e^- e^+ \rightarrow A H\rightarrow ZHH\rightarrow \ell\bar{\ell}\gamma\gamma\gamma\gamma$ where $\ell$ is a muon $\mu$ or an electron $e$ is assumed as the signal process. The $e^- e^+$ collision is assumed to take place at a linear collider operating at $\sqrt{s}=1$ TeV. The signal process has been chosen so that the observation benefits from possible enhancements allowed by the assumed model. Taking $h$ as the SM-like Higgs boson, $\sin(\beta-\alpha)=1$ is assumed so that the $h$-fermion couplings of the Yukawa Lagrangian of Eq. \ref{yukawa1} reduce to the corresponding couplings in the Yukawa Lagrangian of the standard model. As shown in table \ref{BPs}, four benchmark points with different mass hypotheses are assumed.
\begin {table}[h]  
\begin{center}
\begin{tabular}{cccccc} 
& \multicolumn{1}{ c }{\cellcolor{blizzardblue}BP1} & \cellcolor{blizzardblue}BP2 & \cellcolor{blizzardblue}BP3 & \cellcolor{blizzardblue}BP4  \\ \Xhline{3\arrayrulewidth}
\multicolumn{1}{ c  }{\cellcolor{blizzardblue}$m_{h}$} & \multicolumn{4}{ c }{125} \\ 
\multicolumn{1}{ c  }{\cellcolor{blizzardblue}$m_{H}$} & 150 & 200 & 250 & 300 \\
\multicolumn{1}{ c  }{\cellcolor{blizzardblue}$m_{A}$} & 200 & 250 & 300 & 400 \\ 
\multicolumn{1}{ c  }{\cellcolor{blizzardblue}$m_{H^\pm}$} & 200 & 250 & 300 & 400 \\ 
\multicolumn{1}{ c  }{\cellcolor{blizzardblue}$\tan\beta$} & \multicolumn{4}{ c }{40} \\ 
\multicolumn{1}{ c  }{\cellcolor{blizzardblue}$\sin(\beta-\alpha)$} & \multicolumn{4}{ c }{1} \\ \Xhline{3\arrayrulewidth}
\end{tabular}
\caption {Selected benchmark points. \label{BPs}}
\end{center}  
\end {table}
Each benchmark point is simulated and analysed independently. According to the assumed benchmark points, Higgs masses $m_H$ and $m_A$ vary in ranges $150$-$300$ and $200$-$400$ GeV respectively. Also, for all of the benchmark points, $\tan\beta=40$ is assumed for signal to be enhanced as explained in what follows.

The signal process begins by the $e^- e^+$ annihilation into a $Z$ boson. The resultant $Z$ boson experiences the decay $Z\rightarrow AH$ which depends on $\sin(\bma)$ according to table \ref{AHcoupling}, and is thus enhanced in the SM-like limit ($\sin(\beta-\alpha)=1$). In this limit, no $\alpha$-$\beta$ dependence is left for this decay mode and therefore, the signal can benefit from possible enhancements in decays of the Higgs bosons $A$ and $H$, without worrying about any destructive change in the production process amplitude. 
 
\begin{figure}[h!]
  \centering
   \begin{subfigure}[b]{\textwidth}
    \centering 
    \includegraphics[width=0.59\textwidth]{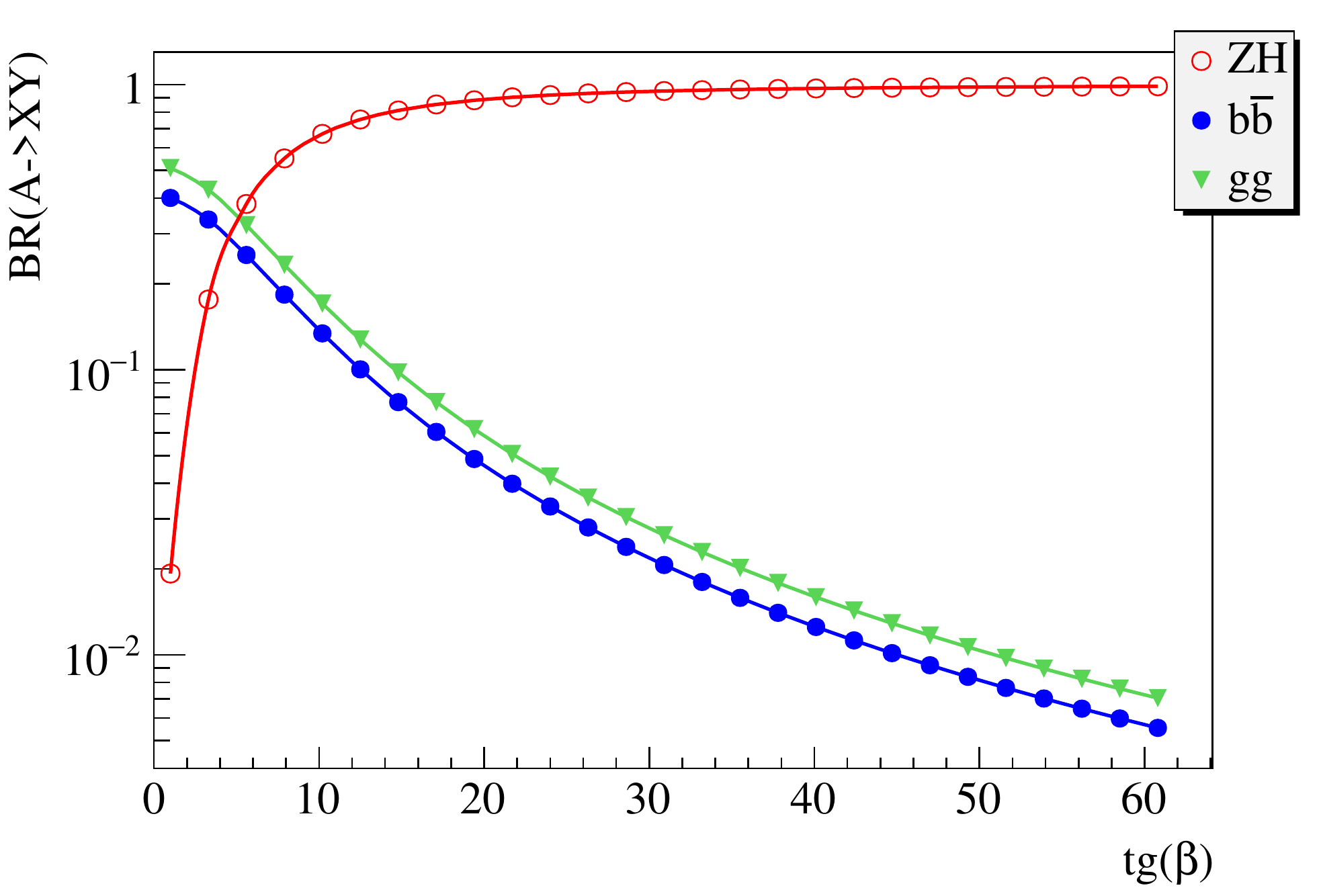}
    \caption{}
    \label{ABRgraph}
    \end{subfigure}
 \quad    
    \begin{subfigure}[b]{\textwidth}
    \centering
    \includegraphics[width=0.59\textwidth]{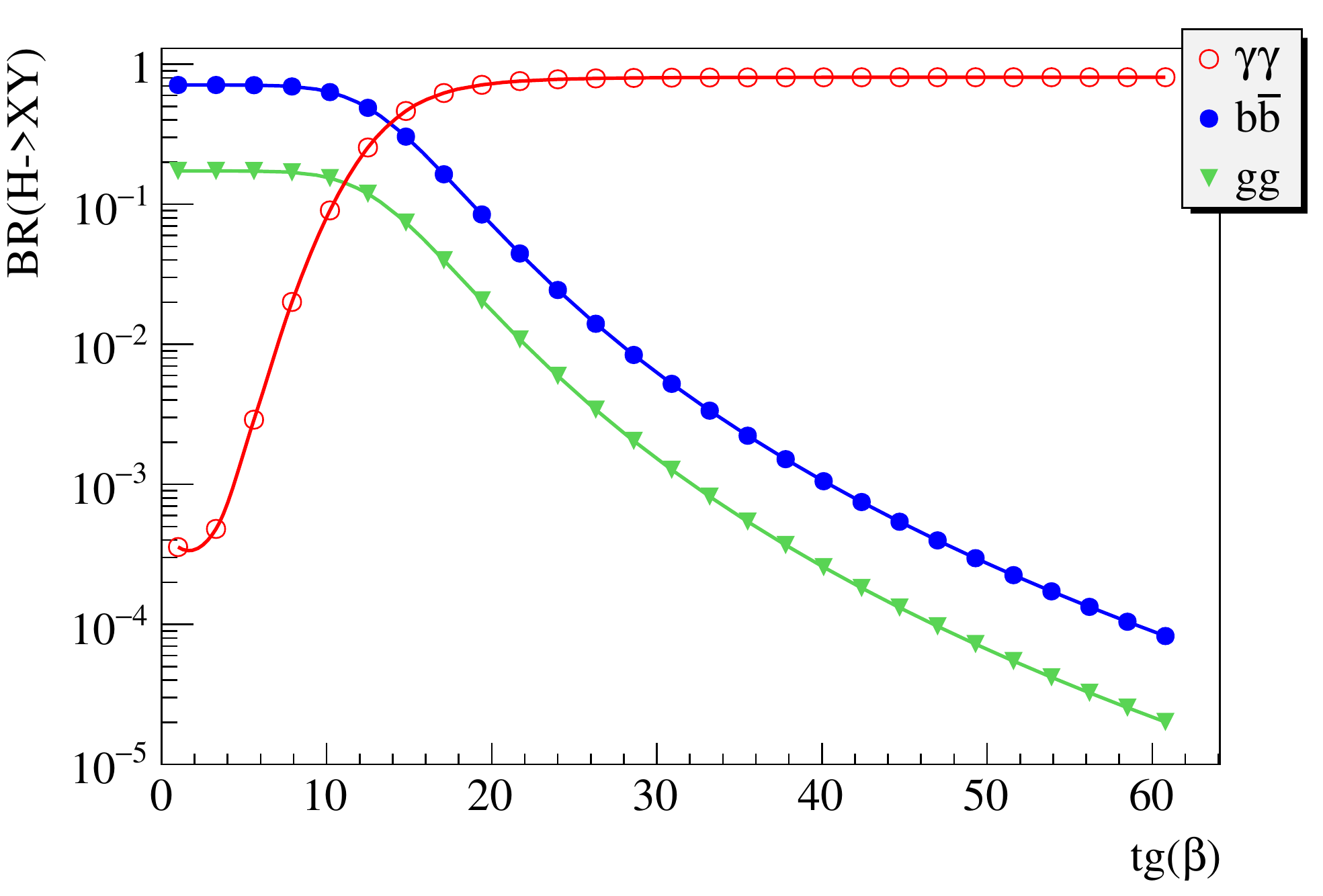}
    \caption{}
    \label{HBRgraph}
    \end{subfigure} 
   \caption{a) $A$ branching ratios into $b\bar{b}$, $gg$ and $ZH$ against $\tan\beta$ assuming BP1, b) $H$ branching ratios into $b\bar{b}$, $gg$ and $\gamma\gamma$ against $\tan\beta$ assuming BP1.}
  \label{HABRgraph}
\end{figure}

\begin{figure}[h!]
  \centering
    \begin{subfigure}[b]{0.2755166666666667\textwidth} 
    \centering   
    \includegraphics[width=\textwidth]{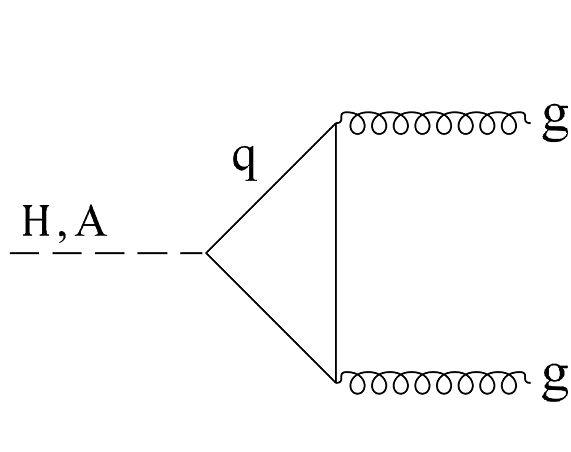} 
    \caption{}
    \label{HAdiagram} 
    \end{subfigure} 
\hspace{15mm}
        \quad    
    \begin{subfigure}[b]{0.61\textwidth}
    \centering
    \includegraphics[width=\textwidth]{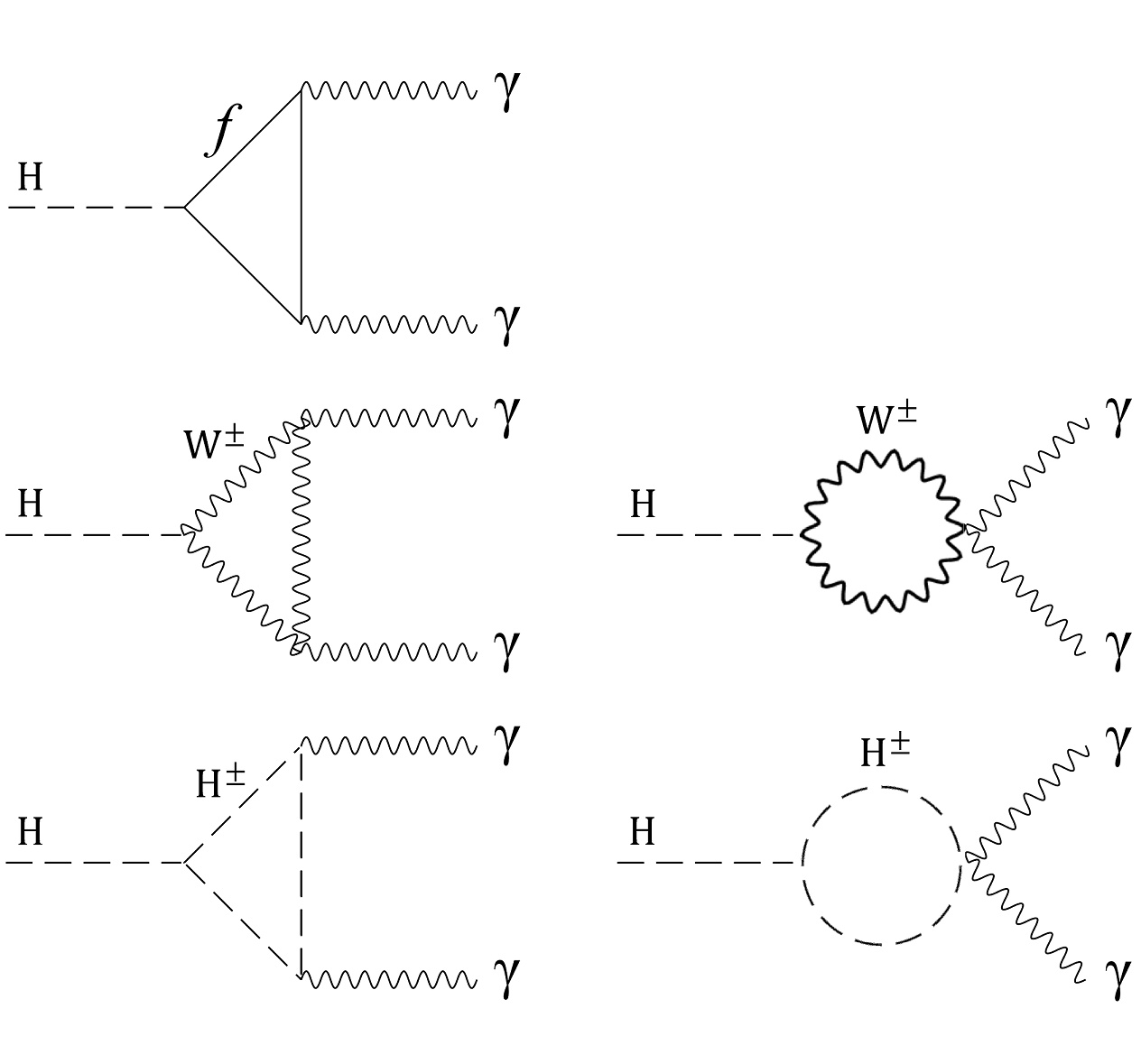}
    \caption{}
    \label{Hdiagram}
    \end{subfigure}
  \caption{Leading order diagrams contributing to processes a) $A/H\rightarrow gg$ and b) $H\rightarrow\gaga$.}
  \label{HHAdiagrams} 
\end{figure}

The produced Higgs $A$ is assumed to decay via mode $A\rightarrow ZH$ which is enhanced for high $\tan\beta$ values in the SM-like limit. Such an enhancement is mainly due to the $\sin(\bma)$ dependence of the $AZH$ coupling (as mentioned earlier) and also the $\cb$ dependence of the $A$ fermionic decays which leads to the suppression of the fermionic decays at high $\tb$ values (see table \ref{AHcoupling}). In Fig. \ref{ABRgraph}, branching ratios of major decay modes of the $A$ Higgs boson is plotted against $\tan\beta$ assuming benchmark point BP1. 
As seen, $A$ branching ratios into $b\bar{b}$ and digluon $gg$ fall and the branching ratio of the $ZH$ mode grows dramatically as $\tan\beta$ increases. Suppression of the digluon mode can be understood as a direct consequence of the suppression of the diquark decays since the digluon decay is a loop-induced decay involving a quark loop as illustrated in Fig. \ref{HAdiagram}. 
For completeness, analytic formulae of the difermion $f\bar{f}$ and digluon decay widths of the Higgs $A$ are given in Eqs. (3.2) to (3.6) \cite{2hdm_HiggsSector2}. $\cot^2\beta$ dependence of the $f\bar{f}$ and $gg$ decay widths is obvious from the corresponding analytic formulae and gives rise to the suppression of these decay modes for large $\tb$ values. It must be noted, as mentioned earlier, that no interaction of type $AVV$ is predicted by the model. Such a feature limits possible decay channels of the $A$ Higgs and therefore can be thought of as another reason behind the large enhancement of the $ZH$ decay mode.

The resultant products $ZHH$ experience the decay modes $Z\rightarrow\ell\bar{\ell}$ and $H\rightarrow\gamma\gamma$ where $\ell\bar{\ell}$ can be a dimuon $\mu^-\mu^+$ or a dielectron $e^-e^+$. Although the leptonic branching ratios of the $Z$ boson is so small ($BR_{Z\rightarrow\mu^-\mu^+\, or\, e^-e^+}\approx0.066$), the leptonic decay mode is chosen to benefit from the clean signature that leptonic events provide in colliders. $H$ decay into a di-photon is of major interest here since not only the di-photon signature is simple and clean, but also the signal can benefit from a large enhancement due to the charged scalar loop contribution to $\gaga$ decay mode for large $\tb$ values. Branching ratios of the major decay modes of the $H$ Higgs boson is plotted against $\tb$ in Fig. \ref{HBRgraph}. As displayed, $b\bar{b}$ and $gg$ decay modes are suppressed and the $\gaga$ decay mode is substantially enhanced for large $\tb$ values. Suppression of the $b\bar{b}$ mode (as well as other fermionic modes) is obviously a consequence of the $\cb$ dependence of the $H$ boson fermionic decay (see table \ref{AHcoupling}). Similarly, suppression of the $gg$ decay mode results from the suppression of the diquark decay since according to Fig. \ref{HAdiagram}, the $gg$ decay involves a quark loop in lowest order just like the $gg$ decay mode of the $A$ Higgs boson. The $\gaga$ decay mode is, however, boosted for large $\tb$ values as explained in the following. Fig. \ref{Hdiagram} shows leading order feynman diagrams contributing to the $\gaga$ decay in a general type-\RN{1} 2HDM. Since the $H$ Higgs boson is gauge-phobic at SM-like limit, the $W^\pm$ loop contribution vanishes and only the three diagrams with fermion and charged Higgs loops contribute to the decay. The fermion loop contribution is suppressed for large $\tb$ values as a result of the $\cb$ dependence of the $Hf\bar{f}$ coupling as seen in table \ref{AHcoupling}. In striking contrast to the fermion loop contribution, the charged Higgs loop contribution is, however, enhanced substantially for large $\tb$ values since the non-vanishing part of the $HH^\pm H^\pm$ coupling at SM-like limit is proportional to $\cot(2\beta)$. The analytic formulae of the $f\bar{f}$, $gg$ and $\gaga$ decay widths of the Higgs $H$ are given in equations \cite{2hdm_HiggsSector2} 
\begin{align} 
&\Gamma(H\to\gamma\gamma) =
\frac{G_F\alpha_\text{EM}^2m_H^3}{128\sqrt2\pi^3}
\left|\sum_fQ_f^2I_f^H(m_f,N_C)+I_{H^\pm}^H\right|^2, \\
&\Gamma(\varphi\to f{\bar f}) = N_C \, \frac{G_Fm_\varphi
m_f^2}{4\sqrt2\pi}\, {\cot^{2}\beta} \times
\begin{cases}
\beta_f^3 \ \ \ \text{ for } \varphi=H\\
\beta_f \ \ \ \text{ for } \varphi=A
\end{cases},\\
&\Gamma(\varphi\to gg) =
\frac{G_F\alpha_S^2m_\varphi^3}{64\sqrt2\pi^3}
\left|\sum_{f=q}I_f^\varphi(m_f,1)\right|^2,
\end{align}
where $N_C=3\,(1)$ for quarks (leptons), $q=u,d,c,s,t,b$ and 
\begin{align}
&\beta_X^{} = \sqrt{1-\frac{4m_X^2}{m_\varphi^2}}\ ,\\
&I_f^\varphi(m_f,\Lambda) = \Lambda\, \frac{8m_f^2}{m_\varphi^2} \,\cb \times
\begin{cases}
1+\beta_f^2\,f\left({4m_f^2}/{m_\varphi^2}\right) 
&\text{ for } \varphi=H\\
\kappa_f\, f\left({4m^2}/{m_\varphi^2}\right) \ \ \ ,\kappa_{f=u(\neq u)}=-1\,(1)   \, \ &\text{ for }
\varphi=A
\end{cases},\\   
&f(x)= 
\begin{cases}
\left[\arcsin\left(\sqrt{1/x}\right)\right]^2&\text{ for } x\ge1\\
-\frac{1}{4}\left[\ln\Big(({1+\sqrt{1-x}})/({1-\sqrt{1-x}})\Big)-i\pi\right]^2
&\text{ for } x<1
\end{cases}.   
\end{align}
According to the analytic formulae, the $f\bar{f}$ and $gg$ decay widths of the Higgs boson $H$ obviously depend on $\cot^2\beta$ which is responsible for the suppression of these decay modes at large $\tb$ values. Also, in the di-photon decay width, the fermion loop contribution $\sum_fQ_f^2I_f^H(m_f,N_C)$ is suppressed because of its dependence on $\cb$. The charged Higgs loop contribution $I_{H^\pm}^H$ which can be found in Ref. \cite{Gunion}, however, depends on $\cot(2\beta)$ which causes the desired enhancement and facilitates reconstruction of the Higgs bosons and thus, to a considerable extent, makes searching for heavy Higgs bosons possible. 

Identifying the products $\ell\bar{\ell}\gaga\gaga$ in the events, $H$ mass $m_H$ is to be computed using the di-photons invariant masses. The invariant mass of the combination of a di-lepton and a di-photon also gives the $A$ mass $m_A$. Reconstructed masses can be extracted from the resultant invariant mass distributions as fully explained in the following sections. 

Signal cross sections listed in table \ref{sXsec} correspond to different benchmark points and are obtained using PYTHIA 8.2.15 \cite{pythia82}.
\begin {table}[h]
\begin{center}
\begin{tabular}{ccccc}
& \multicolumn{1}{ c }{\cellcolor{blizzardblue}BP1} & \cellcolor{blizzardblue}BP2 & \cellcolor{blizzardblue}BP3 & \cellcolor{blizzardblue}BP4 \\ \Xhline{3\arrayrulewidth}
\multicolumn{1}{ c  }{\cellcolor{blizzardblue}Signal cross section [fb]} & 0.470 & 0.273 & 0.167 & 0.088 \\ \Xhline{3\arrayrulewidth}
\end{tabular}
\caption {Cross section of the signal process assuming different benchmark points. \label{sXsec}}
\end{center}
\end {table}
Obviously, signals with heavier Higgs masses have smaller cross sections, and consequently, observing heavier Higgs bosons must be more challenging. Based on the nature of the signal process, major background processes include $W^\pm$ gauge boson pair production, $Z$ gauge boson pair production, top quark pair production and $Z\gamma$ production. Table \ref{bgXsec} presents cross sections corresponding to the background processes, which are obtained by PYTHIA 8.2.15. 
\begin {table}[h]
\begin{center}
\begin{tabular}{ccccc}
& \multicolumn{1}{ c }{\cellcolor{blizzardblue}$T\bar{T}$} & \cellcolor{blizzardblue}$WW$ & \cellcolor{blizzardblue}$ZZ$ & \cellcolor{blizzardblue}$Z\gamma$ \\ \Xhline{3\arrayrulewidth}
\multicolumn{1}{ c  }{\cellcolor{blizzardblue}Cross section [fb]} & 211.1 & 3163 & 234.7 & 4335 \\ \Xhline{3\arrayrulewidth}
\end{tabular}
\caption {Background cross sections. \label{bgXsec}}
\end{center}
\end {table}   

To respect experimental constraints, the deviation of the parameter 
\begin{equation}
\rho=\frac{m_W^2}{(m_Z\cos\theta_W)^{2}}\, ,
\end{equation}
from its standard model value is evaluated to verify whether the deviation is consistent with the experimental constraint \cite{BERTOLINI,DENNER}. The constraint on the $\rho$ parameter in the 2HDM has resulted from the measurement performed at LEP \cite{Yao}. Since it can be shown that the deviation of the $\rho$ parameter from its SM value is negligible if the masses of the Higgs bosons satisfy any of the conditions \cite{drho,Gerard:2007kn}
\begin{equation}
m_A=m_{H^\pm},\,\,\, m_H=m_{H^\pm},
\label{negligibledrho}
\end{equation}
masses of the neutral pseudoscalar ($A$) and charged ($H^\pm$) Higgs bosons are chosen to be equal for all of the assumed benchmark points. By this mass hypothesis, $\rho$ deviation is reduced so that it is consistent with the experimental constraint.

Current experimental limits constrain Higgs bosons masses in the context of the MSSM. As shown in \cite{lep1,lep2,lepexclusion2}, masses of the neutral CP-odd and charged Higgs bosons must meet the conditions $m_A\geq93.4$ GeV and $m_{H^\pm}\geq78.6$ GeV. Moreover, the mass range $m_{A/H}=200-400$ GeV is excluded for $\tb\geq5$ as indicated by the LHC experiments \cite{CMSNeutralHiggs,ATLASNeutralHiggs}. However, the theoretical structure of the MSSM is far different from the type-\RN{1} 2HDM. Not only the Higgs couplings to fermions are different, but also the MSSM possesses less free mass parameters as a result of the imposed symmetry. Therefore, mass hypotheses in the context of the type-\RN{1} 2HDM don't need to be consistent with the experimental constraints obtained based on the MSSM. 

Other than the mentioned limits, the condition $m_{H^\pm}>480$ GeV which results from the flavor physics data \cite{Misiak} in the context of the types \RN{2} and Y of the 2HDM puts a lower limit on the charged Higgs mass. Such a limit, also, doesn't need to be obeyed by our mass hypothesis since the charged Higgs couplings in these types differ considerably from those of the type \RN{1}. More specifically, the $\tb$ dependence of the charged Higgs coupling to quarks in these types considerably affects many of the flavor observables through the diagrams involving the charged Higgs at large values of $\tb$. Such effects are absent in type \RN{1} since the corresponding couplings depend on $\cb$. Hence, in contrast to the types \RN{2} and Y, the type \RN{1} doesn't suffer from such a strong limit on the charged Higgs mass. Finally, it can be concluded that all the assumed benchmark points are safe and consistent with all the theoretical and experimental constraints.  

\section{Event generation, analysis and selection efficiencies} 
For each benchmark point, signal and background events are generated and analysed independently. In order to generate the signal events, model parameters are generated in SLHA (SUSY Les Houches Accord) format using 2HDMC 1.6.3 package and the output files are then passed to PYTHIA 8.2.15 for event generation and further processing including multi-particle interactions, decays, final state showering, etc. Generation of the background events is also performed using PYTHIA 8.2.15. As explained in what follows, signal and background events are analysed and appropriate selection criteria (cuts) are imposed to suppress background events.

\begin{figure}[h!]
  \centering
    \begin{subfigure}[b]{0.59\textwidth}
    \centering
    \includegraphics[width=\textwidth]{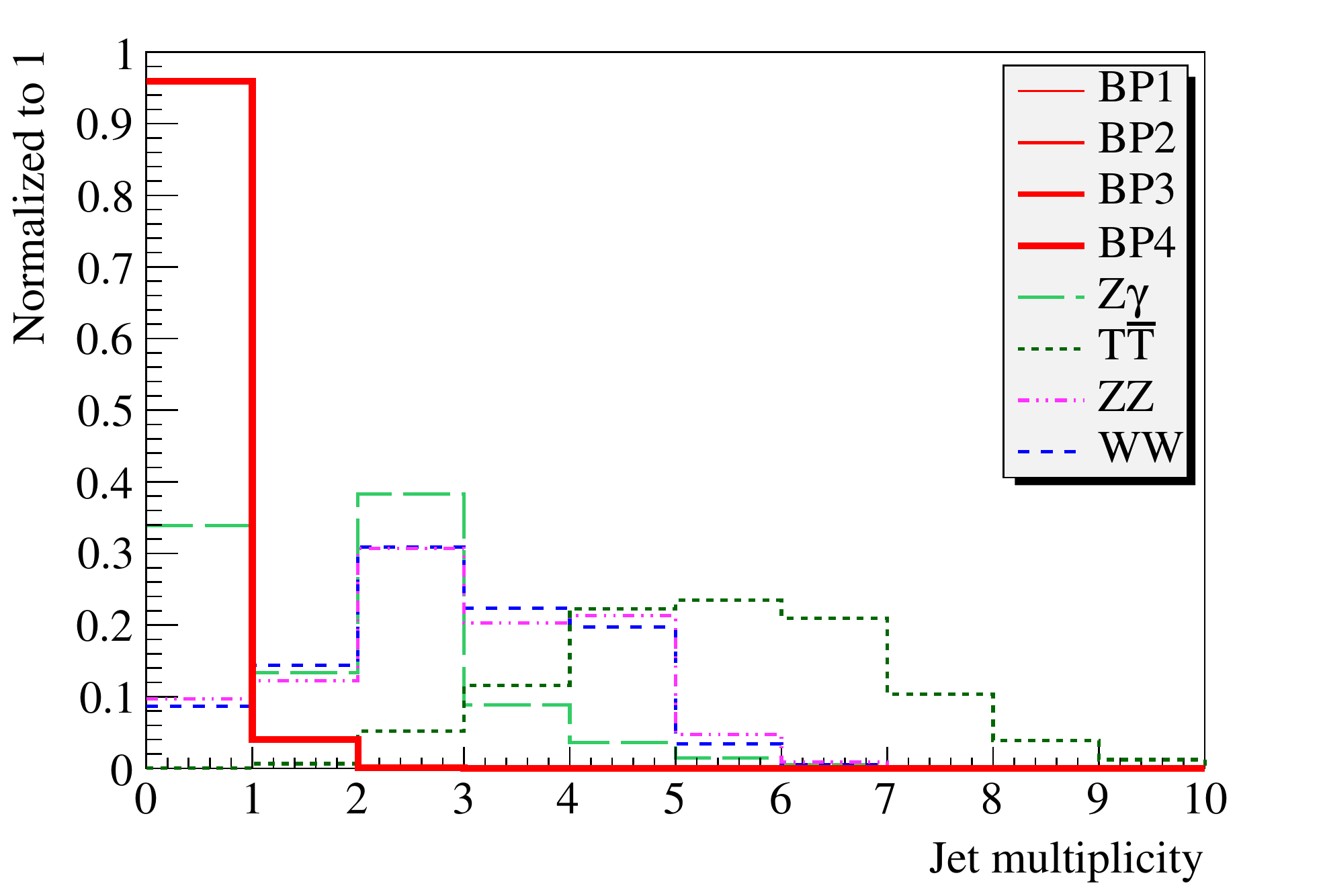}
    \caption{}
    \label{hnjets}
    \end{subfigure}  
        \quad    
    \begin{subfigure}[b]{0.59\textwidth}
    \centering
    \includegraphics[width=\textwidth]{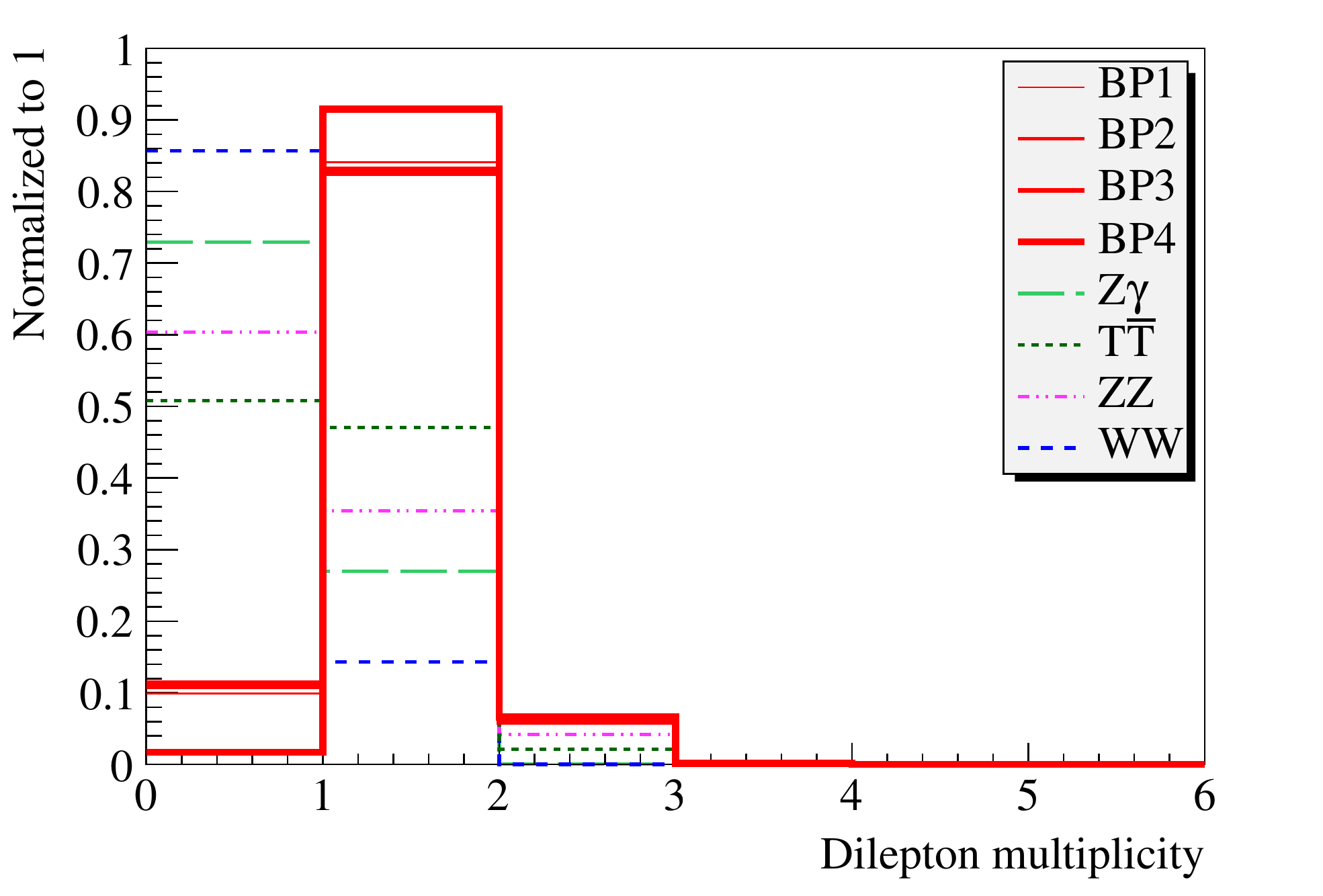}
    \caption{}
    \label{hndileptons}
    \end{subfigure}
  \caption{a) Jet multiplicity and b) di-lepton multiplicity corresponding to different signal (red solid lines) and background (dashed lines) processes. Signal distributions corresponding to different benchmark points (BP1 to BP4) are shown by solid lines with different widths.}
  \label{hnjetshndileptons} 
\end{figure}

Final state constituent particles of the generated events are analysed using FASTJET 3.1.0 \cite{fastjet1,fastjet2} to perform jet reconstruction. Among various sequential recombination clustering algorithms included in this package, anti-$k_t$ algorithm \cite{antikt} with the standard jet cone size $\Delta R=\sqrt{(\Delta\eta)^2+(\Delta\phi)^2}=0.4$, where $\eta=-\textnormal{ln}\tan(\theta/2)$ ($\phi$ and $\theta$ are the azimuthal and polar angles with respect to the beam axis respectively), is used for jet reconstruction. jet energy smearing is applied to jets according to energy resolution $\sigma/E=3.5\, \%$ \cite{cliccdr}. Considering the signal and background processes, the majority of the signal events are expected to have no jets while the background processes are very likely to produce hadronic jets. Hence jet multiplicity distributions of the signal and background events are expected to show significant contrast. Standard jets (reconstructed by the anti-$k_t$ algorithm) which satisfy the conditions 
\begin{equation}
\bm{{p_T}}_{\bm{jet}}\geq10\ GeV,\ \ \  \vert \bm{\eta}_{\bm{jet}} \vert \leq 5,
\label{jetconditions}
\end{equation}
where $p_T$ is the transverse momentum, are counted, and jet multiplicity distributions of Fig. \ref{hnjets} is obtained for signal and background events.
As expected, the distributions differ sharply. The selection cut
\begin{equation}
\bm{N}_{\textbf{\emph{jet}}} \leq 1,
\label{jetsnumbercondition} 
\end{equation}
where $N_{\textbf{\emph{jet}}}$ is the number of jets, is provided by this difference and is imposed on the events.

Identifying lepton content of the surviving events, momentum smearing is applied to leptons according to momentum resolution $\sigma_{p_T}/{p_T^2} = 2 \times 10^{-5}$ GeV$^{-1}$ \cite{cliccdr}, and then only electrons and muons satisfying the threshold conditions
\begin{equation}
{\bm{p_T}}_{\bm{e,\mu}}\geq5\ GeV,\ \ \  \vert \bm{\eta}_{\bm{e,\mu}} \vert \leq 5, 
\label{leptonsconditions}
\end{equation}
are selected. Applying the conditions \ref{leptonsconditions} and counting the number of di-leptons ($e^-e^+$ or $\mu^-\mu^+$), di-lepton multiplicity distributions of Fig. \ref{hndileptons} is obtained. Based on these distributions, the selection cut 
\begin{equation}
\bm{N}_{\bm{\ell\bar{\ell}}} \geq1, 
\label{dileptonscondition}
\end{equation} 
where $N_{\ell\bar{\ell}}$ is the number of di-leptons, is applied. This cut also guarantees the existence of at least one di-lepton which is needed for reconstructing $A$ mass since the $A$ Higgs experiences the gauge-Higgs decay $A\rightarrow ZH$ in the signal process chain.  

\begin{figure}[h!]
  \centering 
    \begin{subfigure}[b]{0.59\textwidth}
    \centering
    \includegraphics[width=\textwidth]{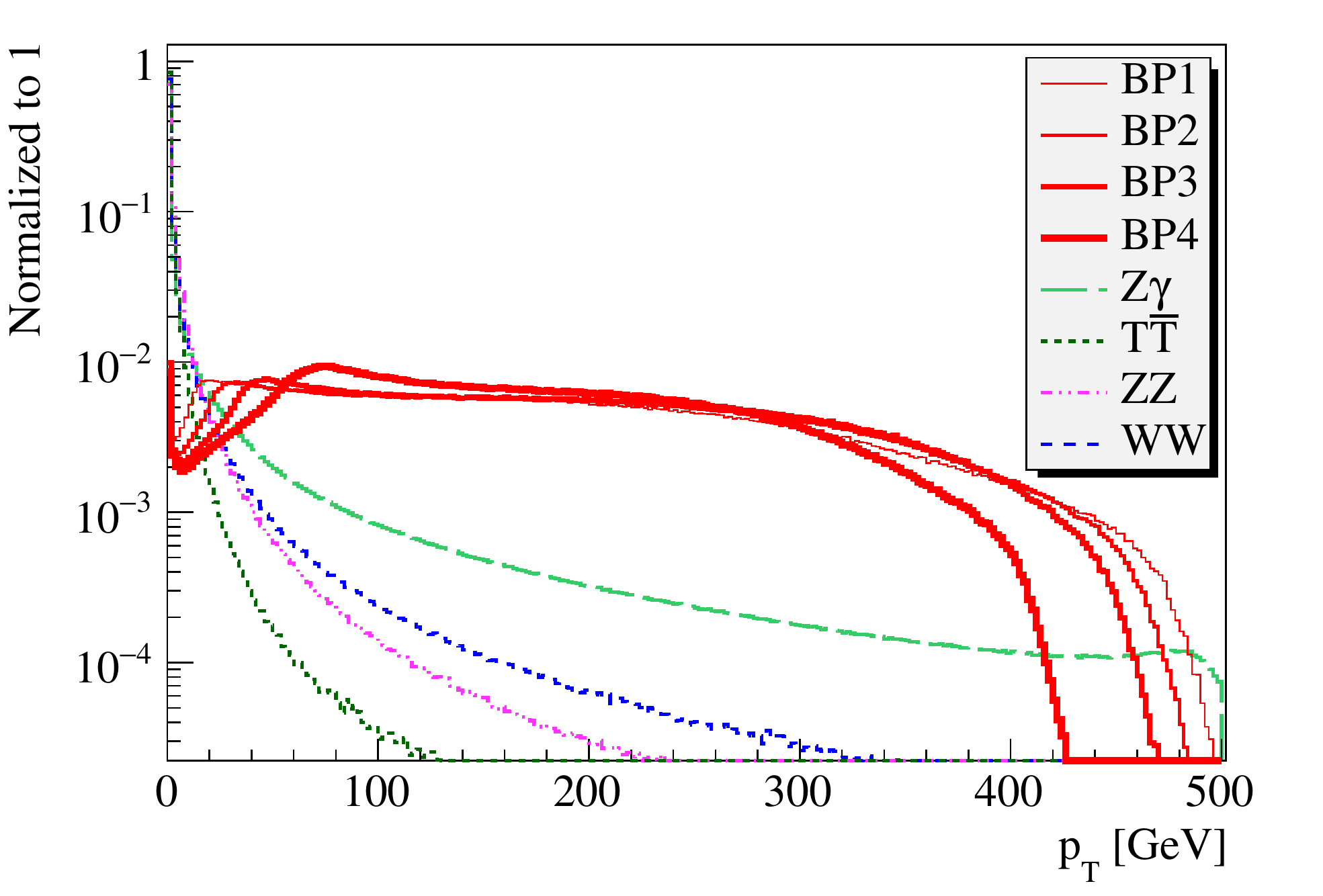}
    \caption{}
    \label{hphotonsptHHIGGS}
    \end{subfigure}
        \quad    
    \begin{subfigure}[b]{0.59\textwidth}
    \centering
    \includegraphics[width=\textwidth]{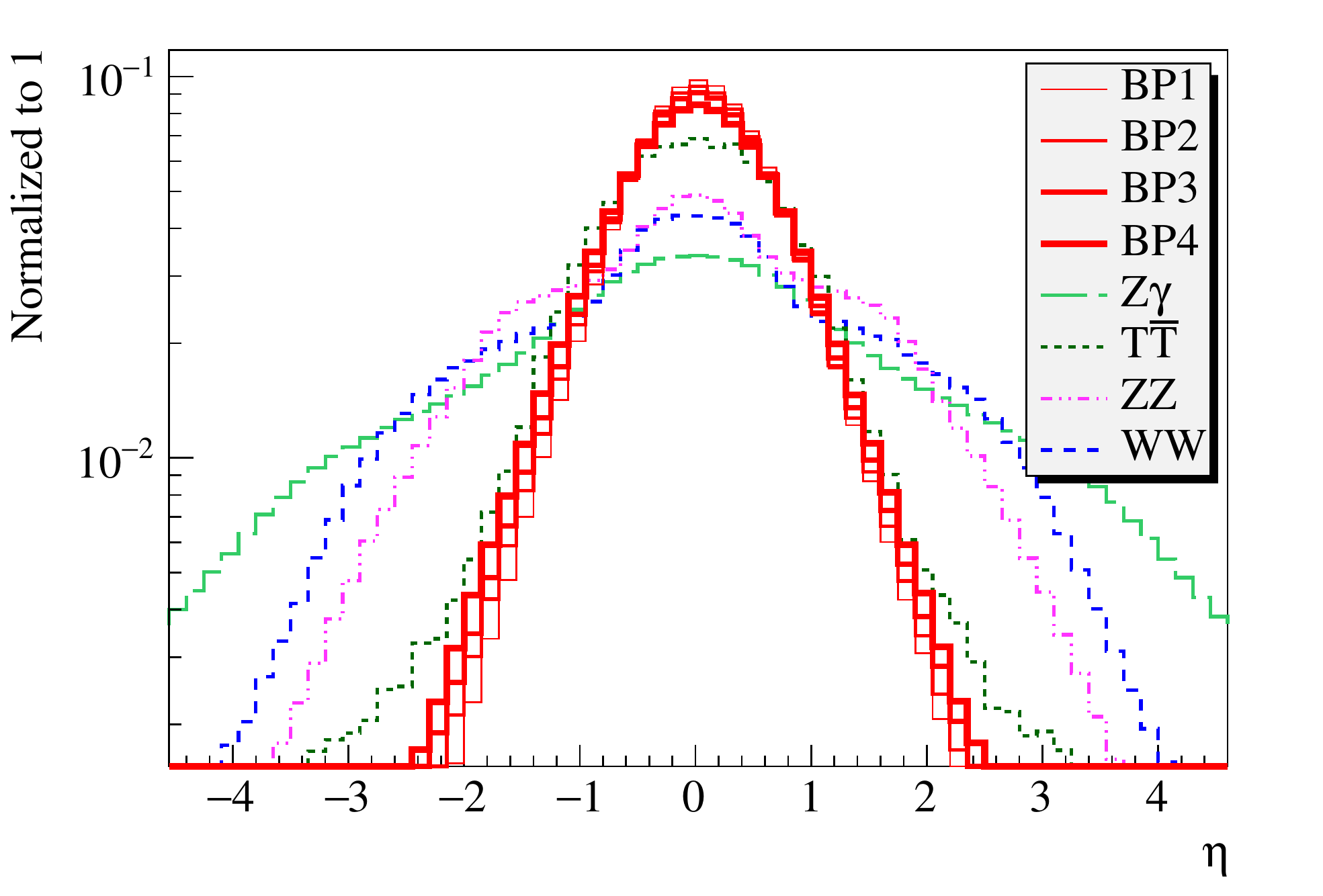}
    \caption{}
    \label{hphotonsetaHHIGGS}
    \end{subfigure} 
  \caption{a) $p_T$ and b) $\eta$ distributions of signal photons originating from $H$ decay (red solid lines) and background photons (dashed lines). $\eta$ distributions are obtained from photons passing the $p_T$ threshold condition \ref{photonsptconditions}.}
  \label{photons}
\end{figure}

Photon content of the surviving events is now identified and photons satisfying kinematic conditions are selected for further analysis and ultimately for $H/A$ reconstruction since $H$ bosons are assumed to decay into di-photons in the signal process chain. Energy smearing is also applied to photons according to energy resolution $\sigma/E=2.7\, \%$ \cite{cliccdr}. In order to determine appropriate kinematic threshold conditions for photons, studying their kinematic properties using information in generator level is useful. Identifying photons produced directly from $H$ decay in signal events using information in generator level and comparing their $p_T$ distribution with $p_T$ distribution of background photons, the plot of Fig. \ref{hphotonsptHHIGGS} is obtained. 
As expected, the average transverse momentum of signal photons resulting from $H$ decay is greater than the average transverse momentum of background photons since background photons originate from relatively light parent particles. The contrast between the patterns and the concentration of the background photons near the zero point suggests a $p_T$ threshold condition harder than the condition applied to leptons. Applying the optimum condition 
\begin{equation}
\bm{{p_T}}_{\bm\gamma}\geq10\ GeV, 
\label{photonsptconditions}
\end{equation}
the $\eta$ distribution of photons passing this condition is obtained as shown in Fig. \ref{hphotonsetaHHIGGS}. The optimum $\eta$ condition
\begin{equation} 
\vert \bm{\eta}_{\bm\gamma} \vert \leq 2.4,
\label{photonsetaconditions}
\end{equation}
which is determined with the help of the Fig. \ref{hphotonsetaHHIGGS}, is also applied to photons. 

\begin{figure}[h]
  \centering
  \includegraphics[width=0.59\textwidth]{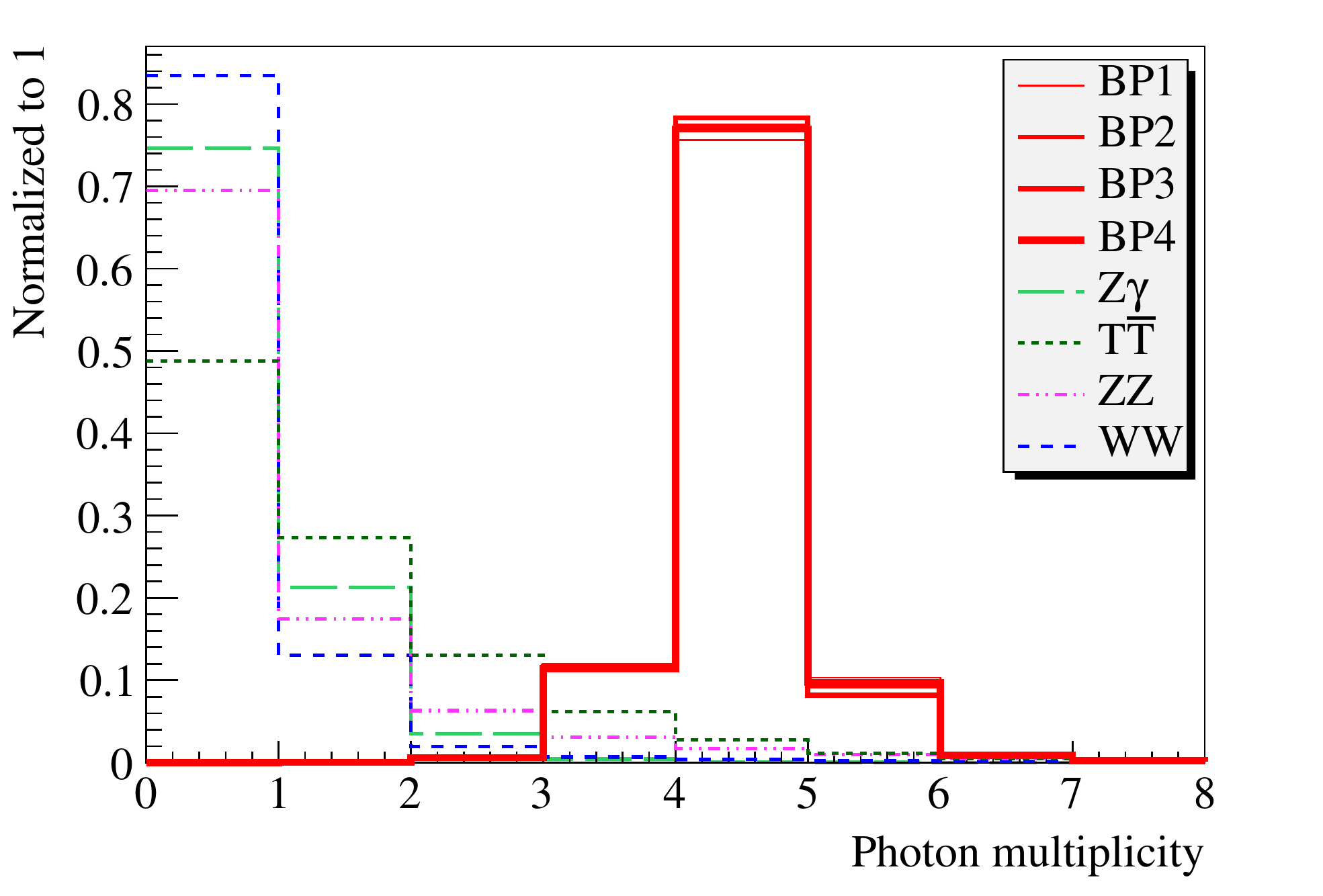}
  \caption{Photon multiplicity distributions of the signal and background events. }
\label{hnphotons}
\end{figure}

Photons surviving the conditions \ref{photonsptconditions} and \ref{photonsetaconditions} are selected for further analysis. They are also counted to obtain photon multiplicity distributions. Obtained distributions are shown in Fig. \ref{hnphotons}. 
As seen, the majority of signal events contain four photons as expected, since both of the $H$ Higgs bosons are assumed to decay into a di-photon. The cut  
\begin{equation} 
{\bm{N}}_{\bm{\gamma}}\geq3,
\label{photonsnumbercondition} 
\end{equation}
based on the sharp contrast between the signal and background distributions of Fig. \ref{hnphotons}, is applied to events.

\begin{figure}[h]
  \centering
  \includegraphics[width=0.59\textwidth]{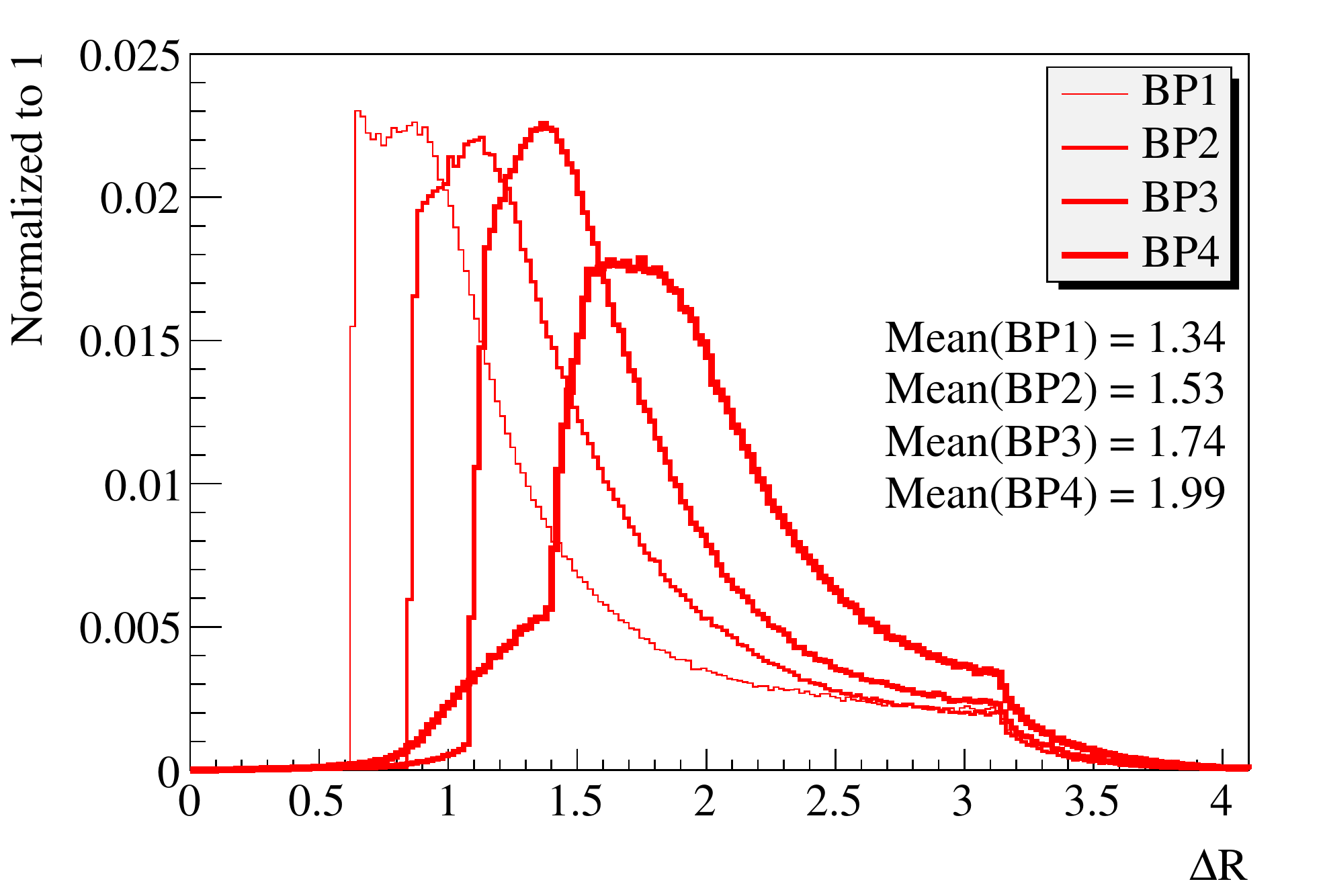}
  \caption{$\Delta R$ distribution corresponding to photon pairs resulting from $H$ decay in signal events surviving the selection cut \ref{photonsnumbercondition} obtained using generator level information.}
\label{hphotonsdRHHIGGS}
\end{figure}

In order to successfully reconstruct the $H$ Higgs boson, true pair(s) of photons must be distinguished. A true photon pair consists of two photons which are decay products of a common parent particle ($H$). In order to find a criterion for true pairs to be distinguished, those signal photons which originate from a common parent $H$ are identified using information in generator level for all signal events surviving the selection cut \ref{photonsnumbercondition}, and then the parameter $\Delta R=\sqrt{(\Delta\eta)^2+(\Delta\phi)^2}$ is computed for all of the identified photon pairs. Here, $\Delta\eta$ ($\Delta\phi$) is the difference in pseudorapidity (azimuthal angle) between the photons of a photon pair. Computing $\Delta R$, the distributions of Fig. \ref{hphotonsdRHHIGGS} is obtained.
As shown in Fig. \ref{hphotonsdRHHIGGS}, the mean values of the parameter $\Delta R$ corresponding to the benchmark points BP1, BP2, BP3 and BP4 are 1.34, 1.53, 1.74 and 1.99 respectively. Taking 1.65, which is the average of the four obtained values, as a criterion for identifying true photon pairs, photon pair selection is performed as follows. In each event, computing the parameter $\Delta R$ for all possible pairs of photons, the pair for which the parameter $\Delta R$ has nearest value to \textbf{1.65} is selected as a true pair. In case the event contains four or more photons, two pairs are selected. The first pair has nearest $\Delta R$ value to 1.65, and the second has second nearest $\Delta R$ value to 1.65.
 
\begin{figure}[h!]
  \centering
    \begin{subfigure}[b]{0.59\textwidth}
    \centering
    \includegraphics[width=\textwidth]{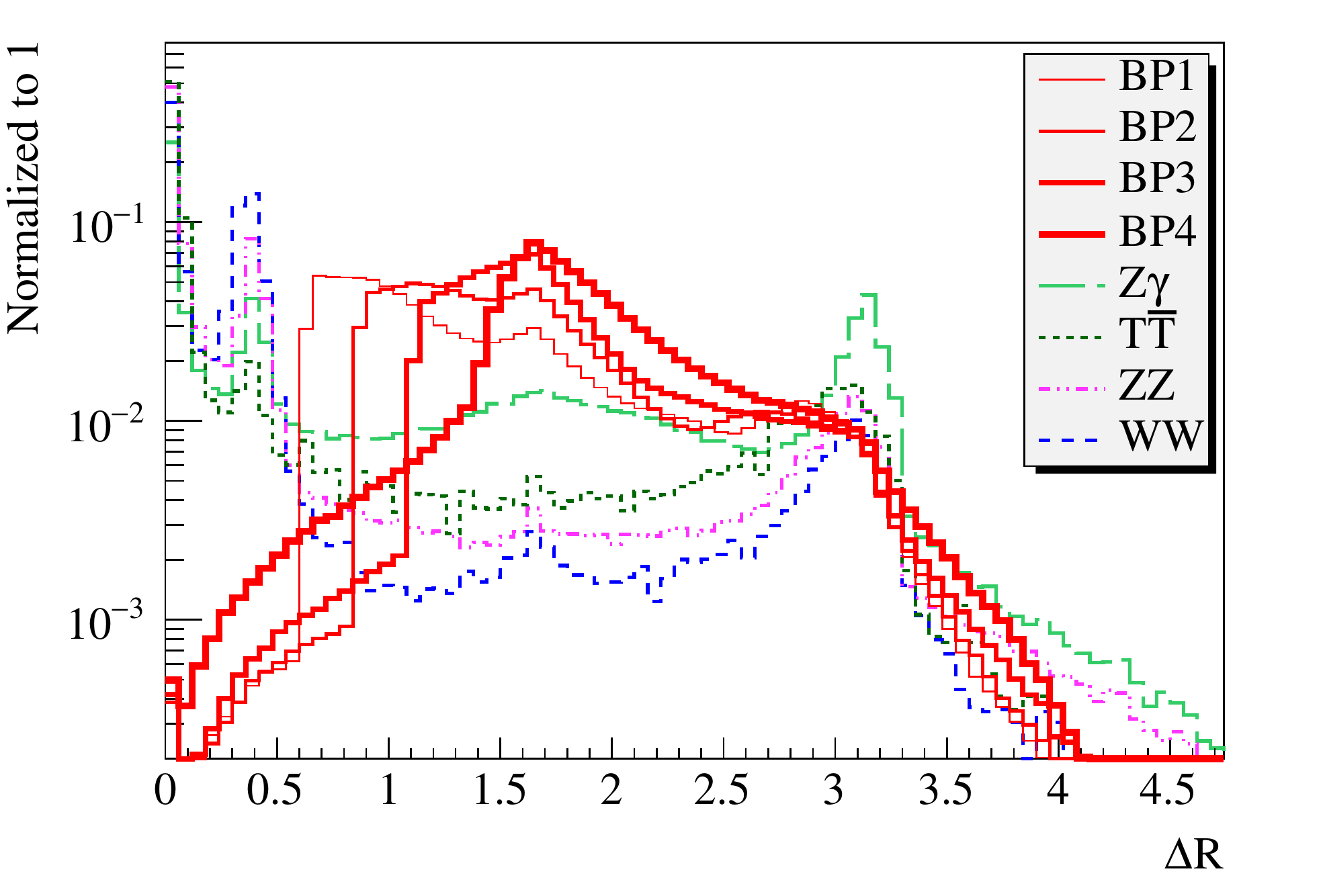}
    \caption{}
    \label{hphotonsdRHcandidate}
    \end{subfigure}
        \quad    
    \begin{subfigure}[b]{0.59\textwidth}
    \centering
    \includegraphics[width=\textwidth]{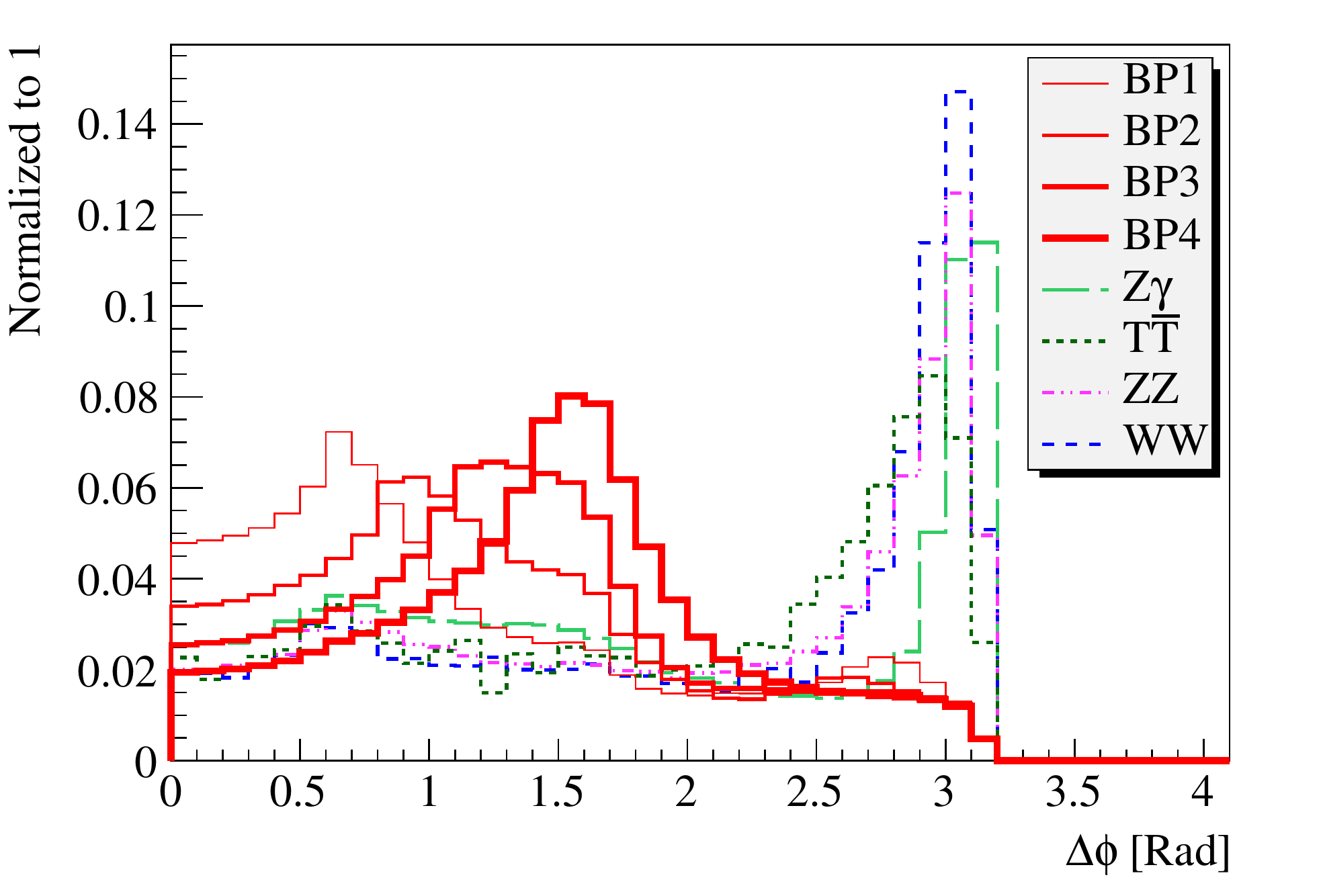}
    \caption{}
    \label{hphotonsdphiHcandidate}
    \end{subfigure} 
  \caption{a) $\Delta R$ and b) $\Delta \phi$ distributions corresponding to selected photon pairs in signal and background events. $\Delta \phi$ distributions are obtained from pairs passing the $\Delta R$ condition introduced in \ref{HcandidatedRdphicondition}.}
  \label{Hcandidate}
\end{figure}

Having selected photon pairs in all events, a condition based on differences in characteristics of the signal and background selected pairs is imposed. Computing the parameter $\Delta R$, this time for the selected photon pairs, the distributions of Fig. \ref{hphotonsdRHcandidate} is obtained.
Based on these distributions, the conditions
\begin{equation} 
0.6 \leq \bm{\Delta R}_{\bm{\gaga}} \leq 4.1, \,\,\,\, \bm{\Delta \phi}_{\bm{\gaga}} \leq 2.9,
\label{HcandidatedRdphicondition}
\end{equation}
are applied to the selected photon pairs. The second condition is suggested by the $\Delta \phi$ distribution corresponding to the pairs passing the first condition, which is shown in Fig. \ref{hphotonsdphiHcandidate}. Events in which none of the selected photon pairs satisfy the conditions \ref{HcandidatedRdphicondition} are ruled out by applying the selection cut
\begin{equation}
\bm{N}_{\bm{\gaga}} \geq1,
\label{diphotonsnumbercondition}
\end{equation}
where $N_{\gaga}$ is the number of pairs satisfying the conditions \ref{HcandidatedRdphicondition}. 

The combination of photons of a photon pair satisfying the conditions \ref{HcandidatedRdphicondition} is considered as the $H$ candidate and the mass distribution obtained from the invariant masses of the photon pairs is used to extract the reconstructed $H$ mass as explained in the following section.

\begin{figure}[h!]
  \centering
    \begin{subfigure}[b]{0.59\textwidth}
    \centering
    \includegraphics[width=\textwidth]{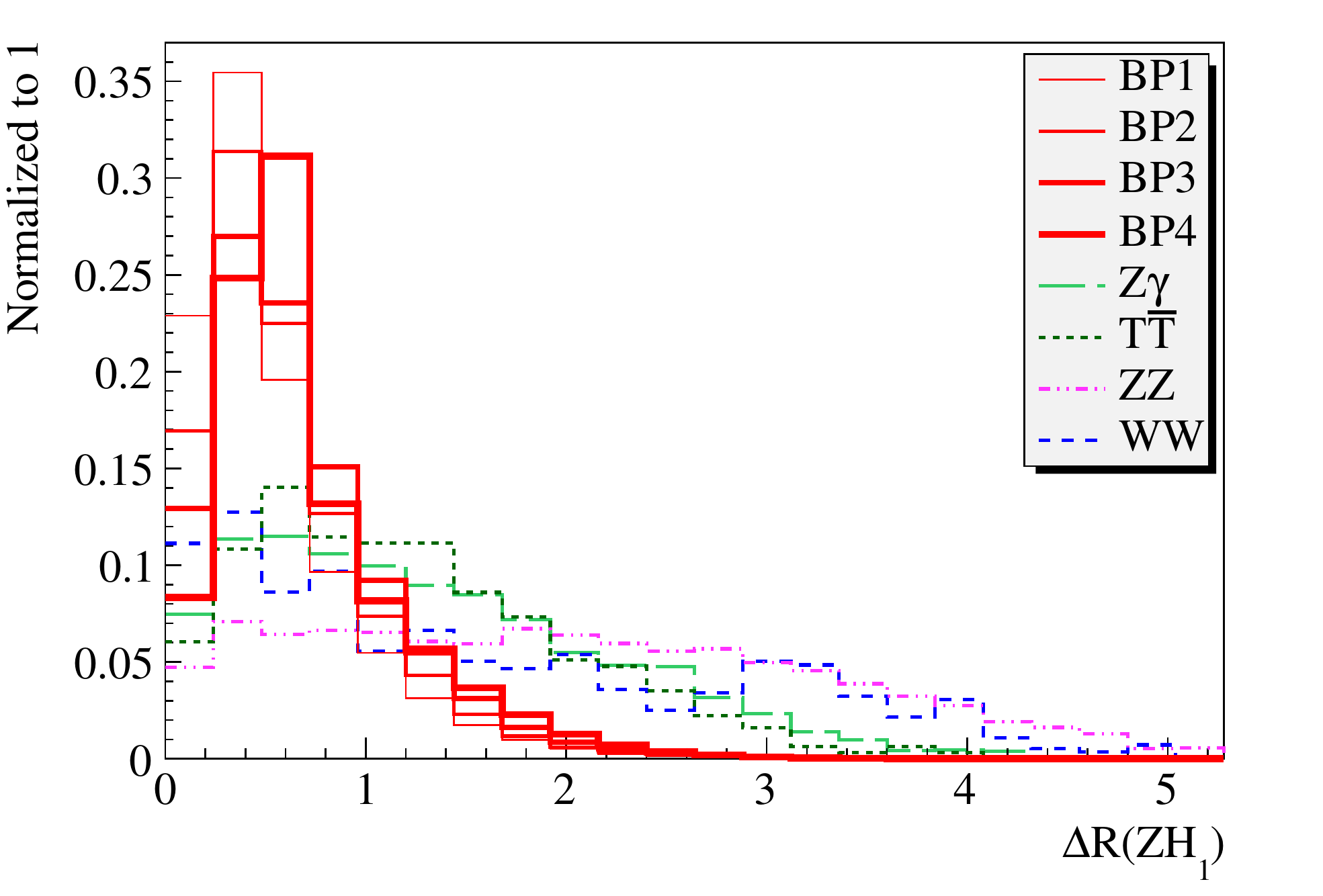}
    \caption{}
    \label{hdRpzHmin}
    \end{subfigure} 
        \quad    
    \begin{subfigure}[b]{0.59\textwidth}
    \centering
    \includegraphics[width=\textwidth]{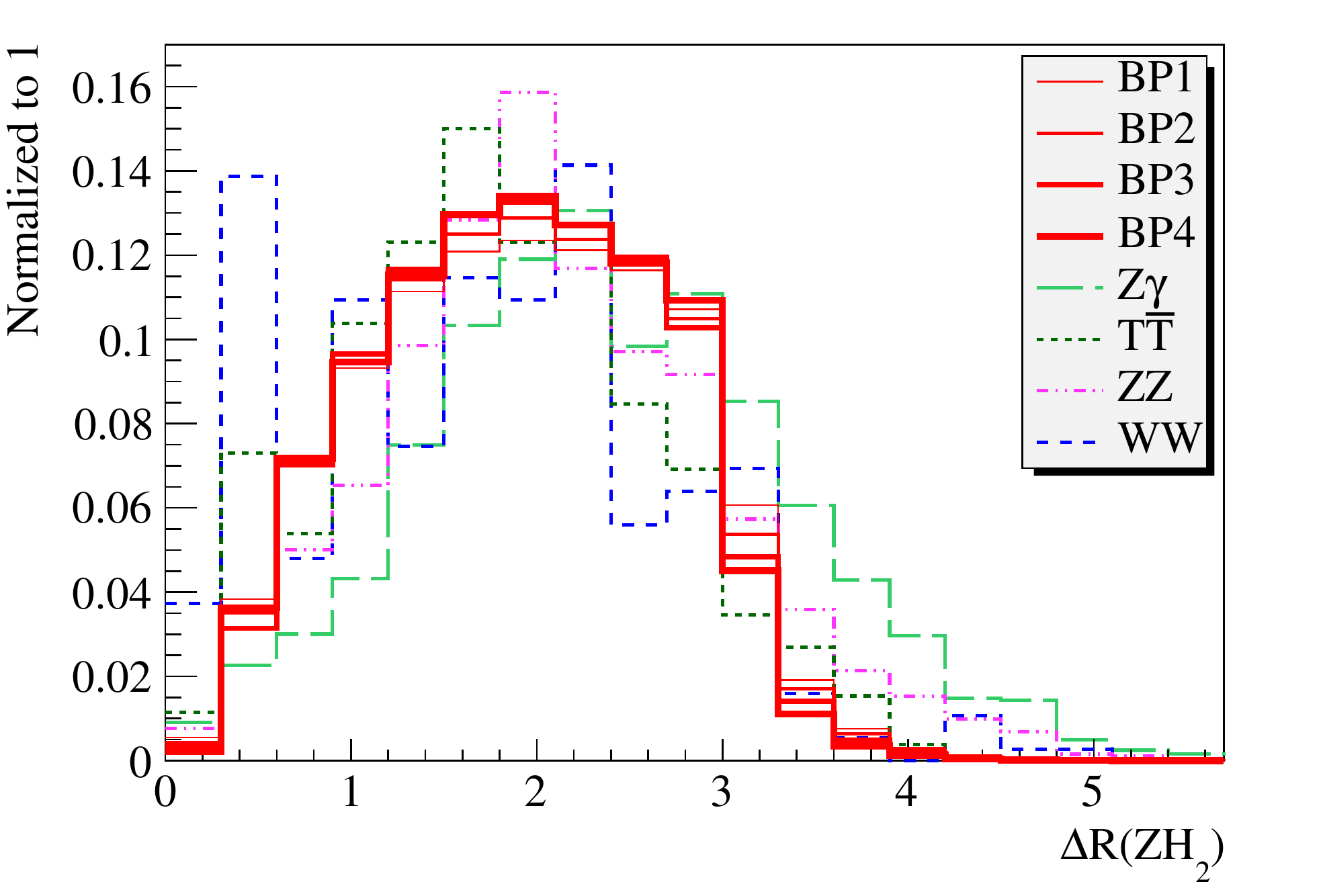}
    \caption{} 
    \label{hdRpzHmaxp}
    \end{subfigure} 
  \caption{$\Delta R$ distributions of a) $ZH_1$ and b) $ZH_2$ pairs for signal and background events. Distributions of figure (b) are obtained from events in which $ZH_1$ satisfies the first condition of \ref{HcandidatedRzhminmax}.}
  \label{Hcandidate}
\end{figure}

Reconstructing the $Z$ boson candidate using the di-lepton $\ell\bar{\ell}$ ($e^-e^+$ or $\mu^-\mu^+$), the distance between the $Z$ and $H$ candidates is measured by computing the parameter $\Delta R$. For events which include two $H$ candidates, $\Delta R$ is computed for both $ZH_1$ and $ZH_2$ pairs. Assuming that the pair with smaller $\Delta R$ value is named $ZH_1$, Fig. \ref{hdRpzHmin} shows the distribution of the $\Delta R$ values corresponding to $ZH_1$ pairs.
Based on these distributions, the conditions 
\begin{equation}
\bm{\Delta R}({\bm{ZH_1}}) \leq 2, \,\,\,\, \bm{\Delta R}({\bm{ZH_2}}) \leq 4.2,
\label{HcandidatedRzhminmax}
\end{equation}
are imposed on $ZH_1$ and $ZH_2$ pairs. The second condition is provided by the distributions of Fig. \ref{hdRpzHmaxp} which are obtained from $\Delta R$ values corresponding to $ZH_2$ pairs in events in which $ZH_1$ passes the first condition introduced in \ref{HcandidatedRzhminmax}. For events with one $H$ candidate, the condition
\begin{equation}
\bm{\Delta R}{({\bm{ZH}})} \leq 0.8, 
\label{HcandidatedRzh}
\end{equation}
is imposed on the only $ZH$ pair. The selection cut 
\begin{equation}  
\bm{N}_{\bm{ZH}} \geq1,
\label{diphotonsnumbercondition}
\end{equation}
is applied to rule out events which lack a $ZH$ pair satisfying the conditions \ref{HcandidatedRzhminmax} (in case of events including two $H$ candidates) or the condition \ref{HcandidatedRzh} (in case of events with one $H$ candidate). 

In events passing this selection cut, the invariant mass of the $ZH$ pair is considered as the $A$ candidate mass and the resultant $A$ candidate mass distribution is used to reconstruct $A$ mass as explained in the following section. In case of events containing two $H$ candidates, the $ZH_1$ combination (which was assumed to have smaller $\Delta R$ value) is taken as the $A$ candidate.

Applying all conditions and selection cuts, signal and background event selection efficiencies are obtained as shown in tables \ref{signalefftab} and \ref{backgroundefftab}. $H$ and $A$ candidate mass distributions are obtained after applying the fourth and the fifth cuts respectively. Total selection efficiencies corresponding to the first four cuts and all the five cuts are also provided in the tables.
\begin{table}[h]
\normalsize
\fontsize{11}{7.2} 
    \begin{center}
         \begin{tabular}{ >{\centering\arraybackslash}m{.72in}  >{\centering\arraybackslash}m{.45in}  >{\centering\arraybackslash}m{.45in} >{\centering\arraybackslash}m{.45in}  >{\centering\arraybackslash}m{.45in}  >{\centering\arraybackslash}m{.45in} } 
  & \cellcolor{blizzardblue}{BP1} & \cellcolor{blizzardblue}{BP2} & \cellcolor{blizzardblue}{BP3} & \cellcolor{blizzardblue}{BP4} \parbox{0pt}{\rule{0pt}{1ex+\baselineskip}}\\ \Xhline{3\arrayrulewidth}
    \cellcolor{blizzardblue}{$N_{jet}\leq1$} & 0.999 & 0.999 & 0.999 & 0.999 \parbox{0pt}{\rule{0pt}{1ex+\baselineskip}}\\ 
   \cellcolor{blizzardblue}{$N_{\ell\bar{\ell}}\geq1$} & 0.901 & 0.893 & 0.886 & 0.983 \parbox{0pt}{\rule{0pt}{1ex+\baselineskip}}\\ 
    \cellcolor{blizzardblue}{$N_{\gamma}\geq3$} & 0.993 & 0.994 & 0.994 & 0.994  \parbox{0pt}{\rule{0pt}{1ex+\baselineskip}}\\ 
\cellcolor{blizzardblue}{$N_{\gaga} \geq1$} & 0.999 & 0.999 & 0.999 & 0.998 \parbox{0pt}{\rule{0pt}{1ex+\baselineskip}}\\ 
    \cellcolor{blizzardblue}{\textbf{Total eff.}} & \textbf{0.894} & \textbf{0.886} & \textbf{0.879} & \textbf{0.974} \parbox{0pt}{\rule{0pt}{1ex+\baselineskip}}\\  \Xhline{3\arrayrulewidth}
   \cellcolor{blizzardblue}{$N_{ZH} \geq1$} & 0.851 & 0.853 & 0.845 & 0.834 \parbox{0pt}{\rule{0pt}{1ex+\baselineskip}}\\ 
    \cellcolor{blizzardblue}{\textbf{Total eff.}} & \textbf{0.761} & \textbf{0.756} & \textbf{0.742} & \textbf{0.812} \parbox{0pt}{\rule{0pt}{1ex+\baselineskip}}\\ \Xhline{3\arrayrulewidth}
        \end{tabular}
\caption{Signal selection efficiencies assuming different benchmark points. \label{signalefftab}}
  \end{center}
\end{table}
\begin{table}[h] 
\normalsize
\fontsize{11}{7.2} 
    \begin{center} 
         \begin{tabular}{ >{\centering\arraybackslash}m{.72in}  >{\centering\arraybackslash}m{.62in}  >{\centering\arraybackslash}m{.62in} >{\centering\arraybackslash}m{.62in} >{\centering\arraybackslash}m{.62in}  >{\centering\arraybackslash}m{.62in} }
& \cellcolor{blizzardblue}{$T\bar{T}$} & \cellcolor{blizzardblue}{$WW$} & \cellcolor{blizzardblue}{$ZZ$} & \cellcolor{blizzardblue}{$Z\gamma$ } \parbox{0pt}{\rule{0pt}{1ex+\baselineskip}}\\ \Xhline{3\arrayrulewidth}
    \cellcolor{blizzardblue}{$N_{jet}\leq1$} & 0.00674 & 0.23028 & 0.21905 & 0.47020  \parbox{0pt}{\rule{0pt}{1ex+\baselineskip}}\\ 
    \cellcolor{blizzardblue}{$N_{\ell\bar{\ell}}\geq1$} & 0.49210 & 0.14341 & 0.39633 & 0.27054 \parbox{0pt}{\rule{0pt}{1ex+\baselineskip}}\\ 
    \cellcolor{blizzardblue}{$N_{\gamma}\geq3$} & 0.10855 & 0.01606 & 0.067776 & 0.00576 \parbox{0pt}{\rule{0pt}{1ex+\baselineskip}}\\ 
    \cellcolor{blizzardblue}{$N_{\gaga} \geq1$} & 0.30251 & 0.13776 & 0.20340 & 0.48406  \parbox{0pt}{\rule{0pt}{1ex+\baselineskip}}\\ 
    \cellcolor{blizzardblue}{\textbf{Total eff.}} & \textbf{0.00011} & \textbf{0.00007} & \textbf{0.00120} & \textbf{0.00035} \parbox{0pt}{\rule{0pt}{1ex+\baselineskip}}\\ \Xhline{3\arrayrulewidth}
   \cellcolor{blizzardblue}{$N_{ZH} \geq1$} & 0.27136 & 0.24938 & 0.19613 & 0.19908 \parbox{0pt}{\rule{0pt}{1ex+\baselineskip}}\\ 
   \cellcolor{blizzardblue}{\textbf{Total eff.}} & \textbf{0.00003} & \textbf{0.00002} & \textbf{0.00023} &  \textbf{0.00007} \parbox{0pt}{\rule{0pt}{1ex+\baselineskip}}\\ \Xhline{3\arrayrulewidth}
  \end{tabular}
\caption{Background selection efficiencies. \label{backgroundefftab}}
  \end{center}
\end{table}

\section{Reconstruction of the Higgs bosons $H$ and $A$ }
Selected di-photons in events surviving the first four selection cuts are considered as the $H$ decay products and thus their invariant masses are used to obtain the $H$ candidate mass distribution. In the signal process, the $A$ Higgs experiences the decay process $A\rightarrow ZH$. Accordingly, in events which pass all of the five cuts, the combination of the identified di-lepton and di-photon is considered as the $A$ candidate. In events with two $H$ candidates, the $ZH$ combination for which the parameter $\Delta R$ has smaller value is taken as the $A$ candidate. 

\begin{figure}[h!]
  \centering
    \begin{subfigure}[b]{0.59\textwidth}
    \centering
    \includegraphics[width=\textwidth]{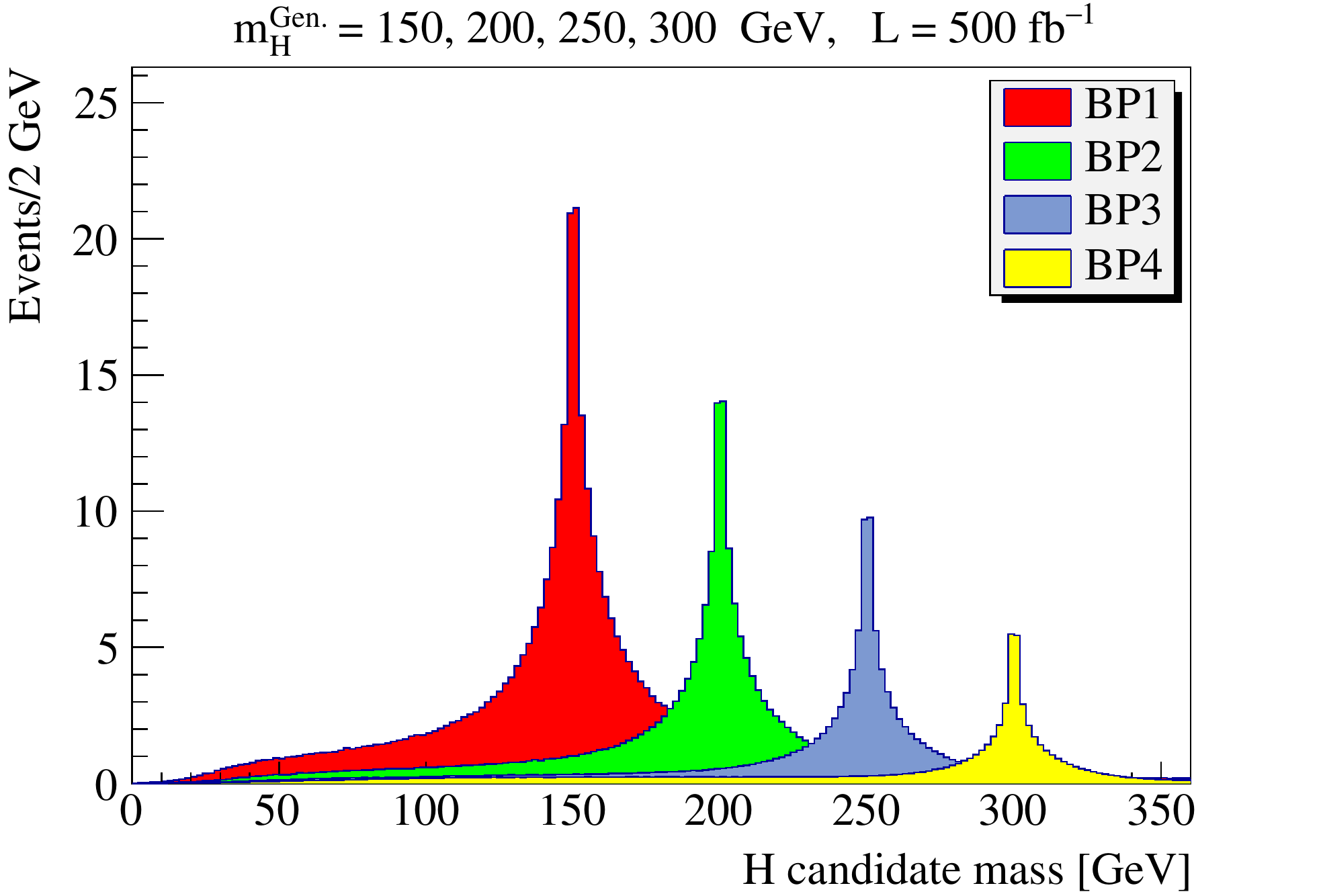} 
    \caption{}
    \label{H-allmHMasses-S}
    \end{subfigure}
    \medskip 
    \quad    
    \begin{subfigure}[b]{0.59\textwidth}
    \centering
    \includegraphics[width=\textwidth]{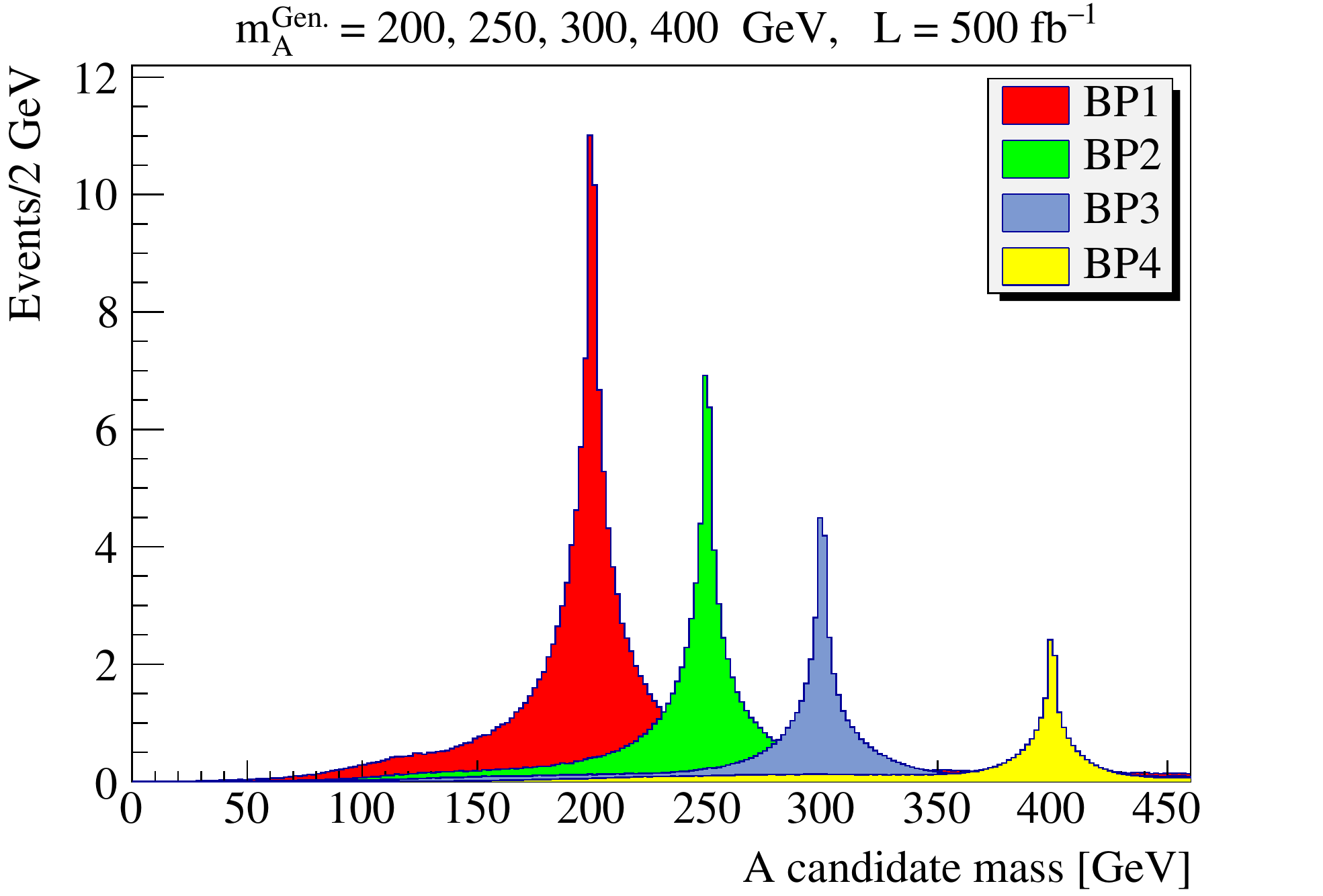}
    \caption{}
    \label{A-allmHMasses-S}
    \end{subfigure} 
  \caption{a) $H$ and b) $A$ candidate mass distributions of the signal events corresponding to different benchmark points.}
  \label{HA-allmHmasses-S}
\end{figure}

Reconstructing the Higgs bosons $H$ and $A$ for signal events as described, the plots of Fig. \ref{HA-allmHmasses-S} are obtained. 
Normalization is based on $L\times\sigma\times\epsilon$, where $L$ is the integrated luminosity which is set to be 500 $fb^{-1}$, $\sigma$ is the signal cross section (given in table \ref{sXsec}) and $\epsilon$ is the total efficiency. Total efficiencies used for $A$ distributions are taken from the last row of table \ref{signalefftab}. Total efficiencies corresponding to the first four cuts provided in table \ref{signalefftab} are not used for normalizing $H$ distributions since the number of identified $H$ candidates in signal events differs from event to event. Total efficiencies 
\begin{equation}
\bm{\epsilon_{BP1}}=0.78,\ \ \bm{\epsilon_{BP2}}=0.80,\ \ \bm{\epsilon_{BP3}}=0.80,\ \ \bm{\epsilon_{BP4}}=0.88,
\label{signalnormaleff}
\end{equation} 
multiplied by 2, are used for normalizing $H$ signal distributions. Multiplication by 2 is because of the fact that, according to the signal process, two $H$ bosons are produced in each signal event. Benchmark points BP1, BP2, BP3, BP4 correspond to generated masses ${m_H}=150, 200, 250, 300$ and ${m_A}=200, 250, 300, 400$ GeV respectively. As seen in Fig. \ref{HA-allmHmasses-S}, signal distributions show sharp peaks almost at the generated masses. 

\begin{figure}[h!]
  \centering
    \begin{subfigure}[b]{0.59\textwidth}
    \centering
    \includegraphics[width=\textwidth]{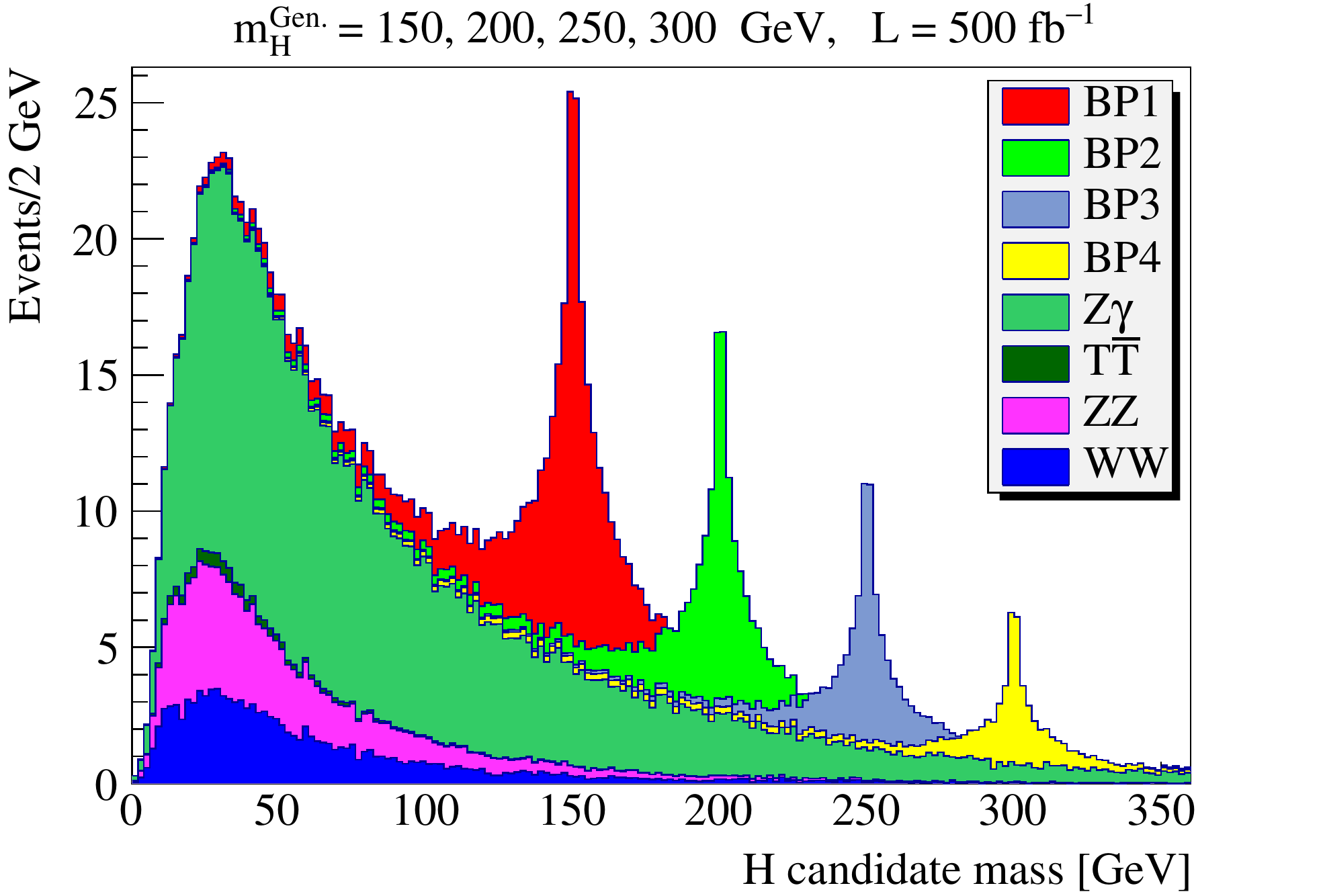}
    \caption{}
    \label{H-allmHmasses} 
    \end{subfigure}
    \medskip
    \quad    
    \begin{subfigure}[b]{0.59\textwidth}
    \centering
    \includegraphics[width=\textwidth]{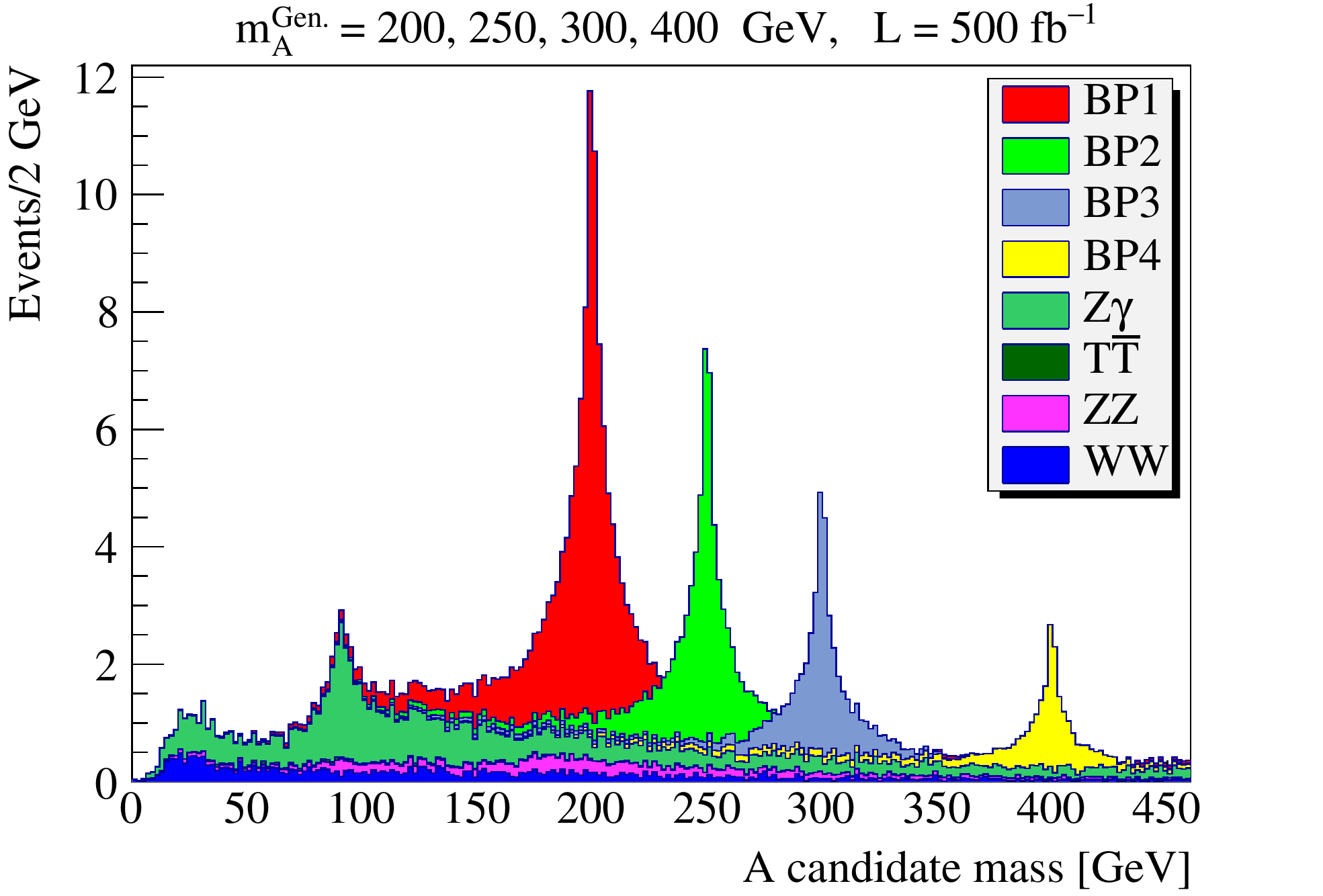}
    \caption{}
    \label{A-allmHmasses}
    \end{subfigure} 
      \caption{Signal plus background candidate mass distributions for a) $H$ and b) $A$ Higgs bosons. Distributions corresponding to different background processes are shown separately.}
  \label{HA-allmHmasses}
\end{figure}

Adding Higgs candidate mass distributions corresponding to the signal and background events together, plots of Fig. \ref{HA-allmHmasses} are obtained. 
According to Fig. \ref{HA-allmHmasses}, signal peaks corresponding to the four assumed benchmark points can be seen on top of the background distributions, and are well distinguished from the background for both of the Higgs bosons. Total efficiencies 
\begin{equation}
\bm{\epsilon_{T\bar{T}}}=0.00012,\ \ \bm{\epsilon_{WW}}=0.00007,\ \ \bm{\epsilon_{ZZ}}=0.00126,\ \ \bm{\epsilon_{Z\gamma}}=0.00034,
\label{backgroundnormaleff}
\end{equation}
are used for normalizing background contributions to $H$ candidate mass distribution. Normalization of background contributions to $A$ candidate mass distribution is done using the total efficiencies provided in the last row of table \ref{backgroundefftab}. As seen, the $T\bar{T}$ background process makes almost no contribution to the total background distribution. This is mostly due to the relatively small cross section (see table \ref{bgXsec}) and also the relatively small efficiency corresponding to the number of jets $N_{jets}$ selection cut as seen in table \ref{backgroundefftab}. Smallness of the efficiency of this cut was expected because of the relatively large number of jets produced by top decay in this process. It is also seen that the $Z\gamma$ contribution is dominant, which is because of the relatively large cross section of this process. Both of $H$ and $A$ mass distributions show sharp peaks almost at the generated masses. Apart from the peaks due to the signal, the $A$ candidate mass distribution of Fig. \ref{A-allmHmasses} shows a small peak mainly due to the $Z\gamma$ process and centred almost at the $Z$ boson mass ($\approx 90$ GeV). This was also expected since di-photons in $Z\gamma$ events are mostly low energy. As a result, invariant mass of the combination $Z\gamma\gamma$ tends to be close to the $Z$ boson mass.

\begin{figure}[h!]
  \centering 
  \bigskip
    \begin{subfigure}[b]{0.8\textwidth} 
    \centering 
     \includegraphics[width=\textwidth]{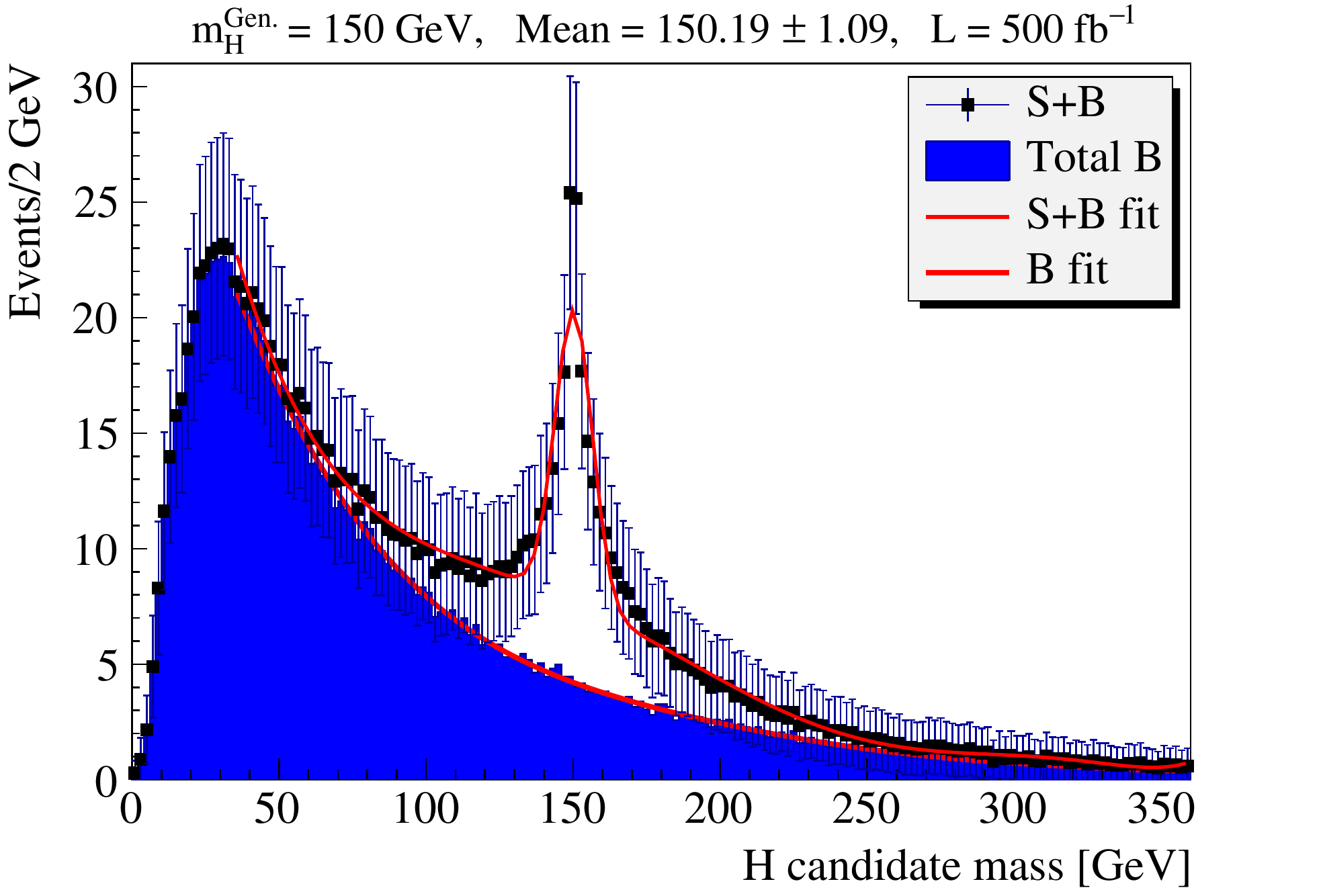}
    \caption{}
    \label{H150Poly.pdf}
    \end{subfigure}  
    \bigskip 
    \quad
    \begin{subfigure}[b]{0.8\textwidth} 
    \centering 
    \includegraphics[width=\textwidth]{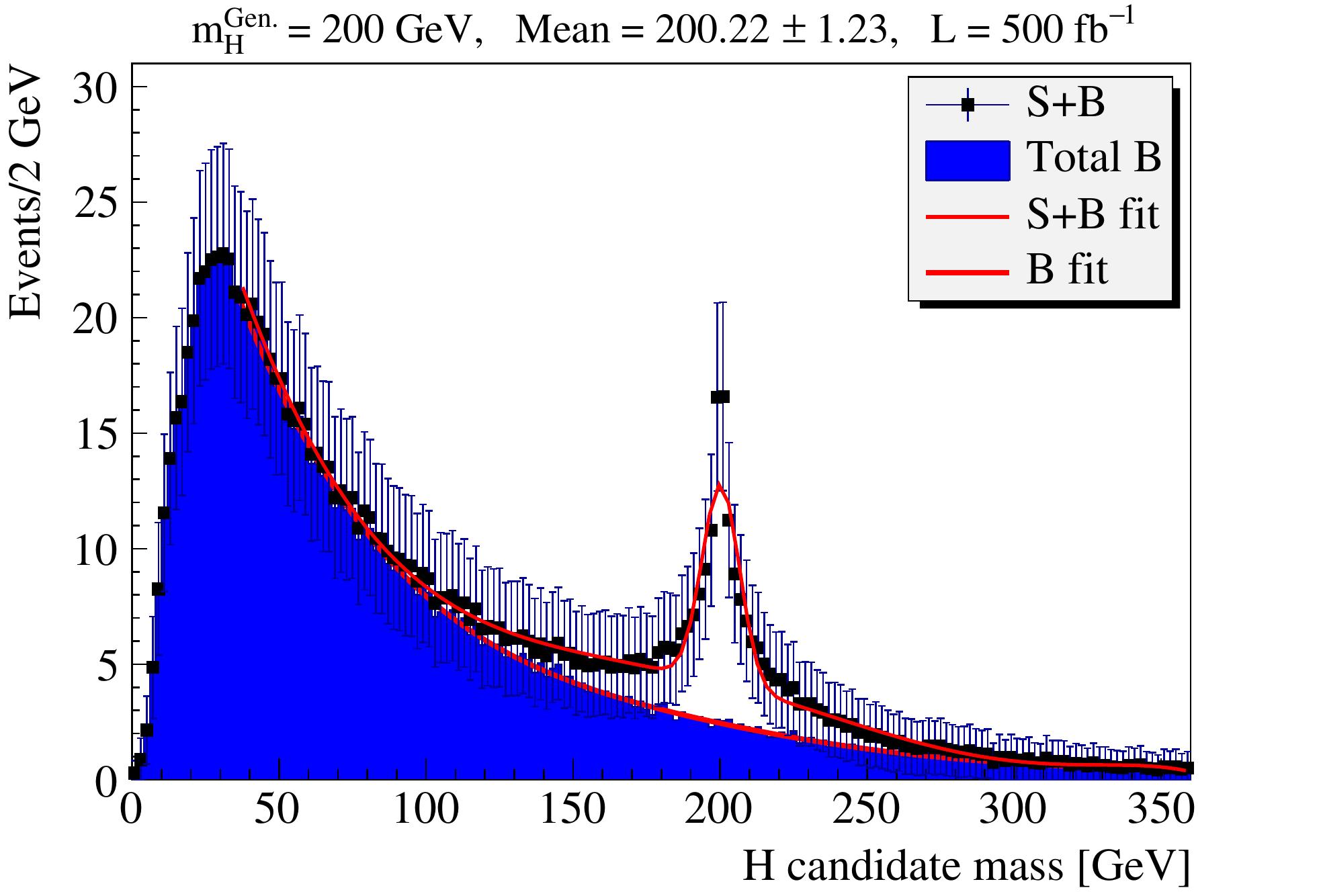}
    \caption{}
    \label{H200Poly.pdf} 
    \end{subfigure} 
\caption{Fitting results of the $H$ candidate mass distributions corresponding to the benchmark points a) BP1 and b) BP2. A polynomial function is fitted to the total background distribution. Statistical errors of the simulated data and the gaussian function ``mean'' parameter values are also shown.}
  \label{H1fit}
\end{figure}

\begin{figure}[h!]
   \bigskip
  \centering  
    \begin{subfigure}[b]{0.8\textwidth}
    \centering
    \includegraphics[width=\textwidth]{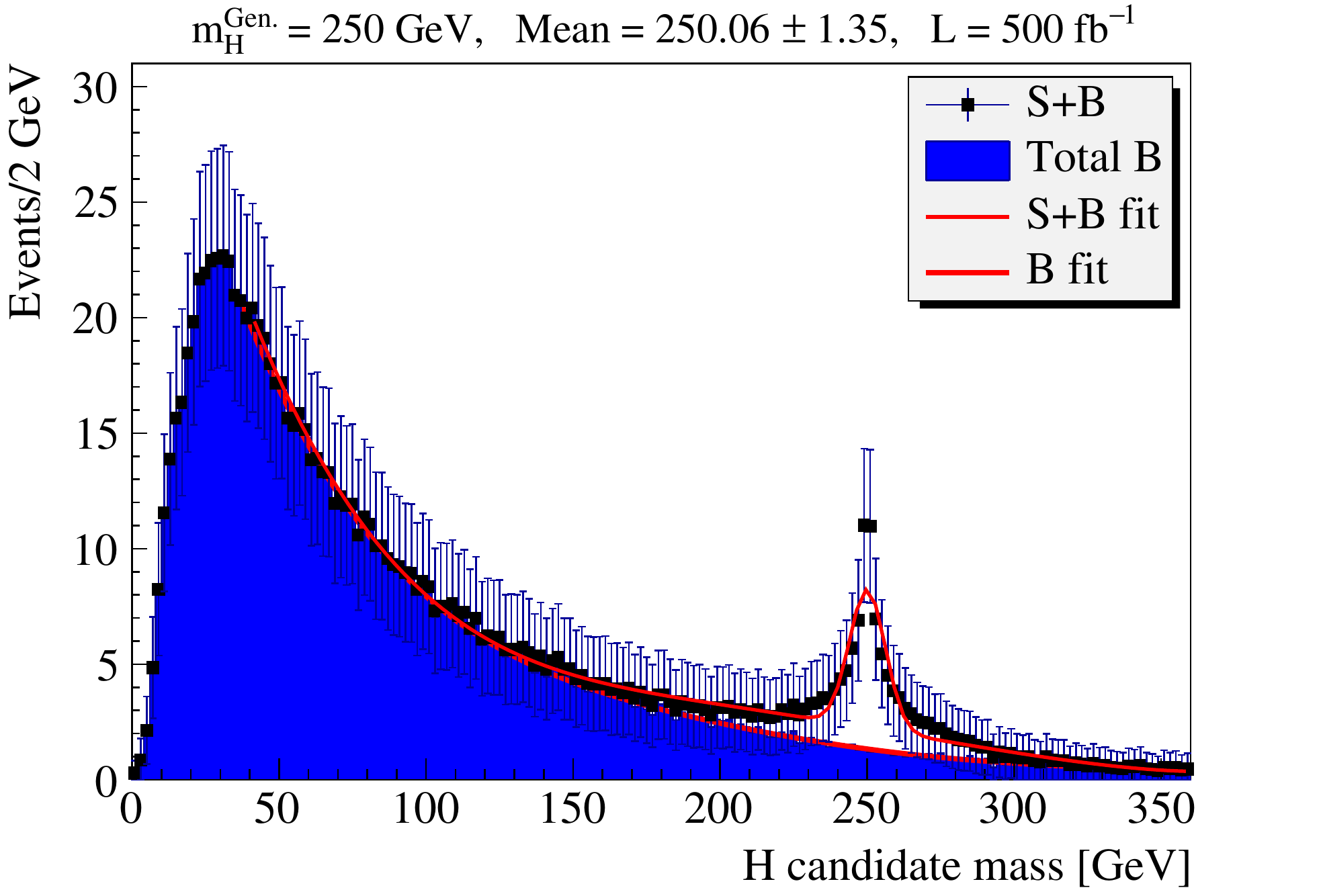}
    \caption{}
    \label{H250Poly.pdf}
    \end{subfigure} 
    \bigskip
    \quad    
    \begin{subfigure}[b]{0.8\textwidth}
    \centering
    \includegraphics[width=\textwidth]{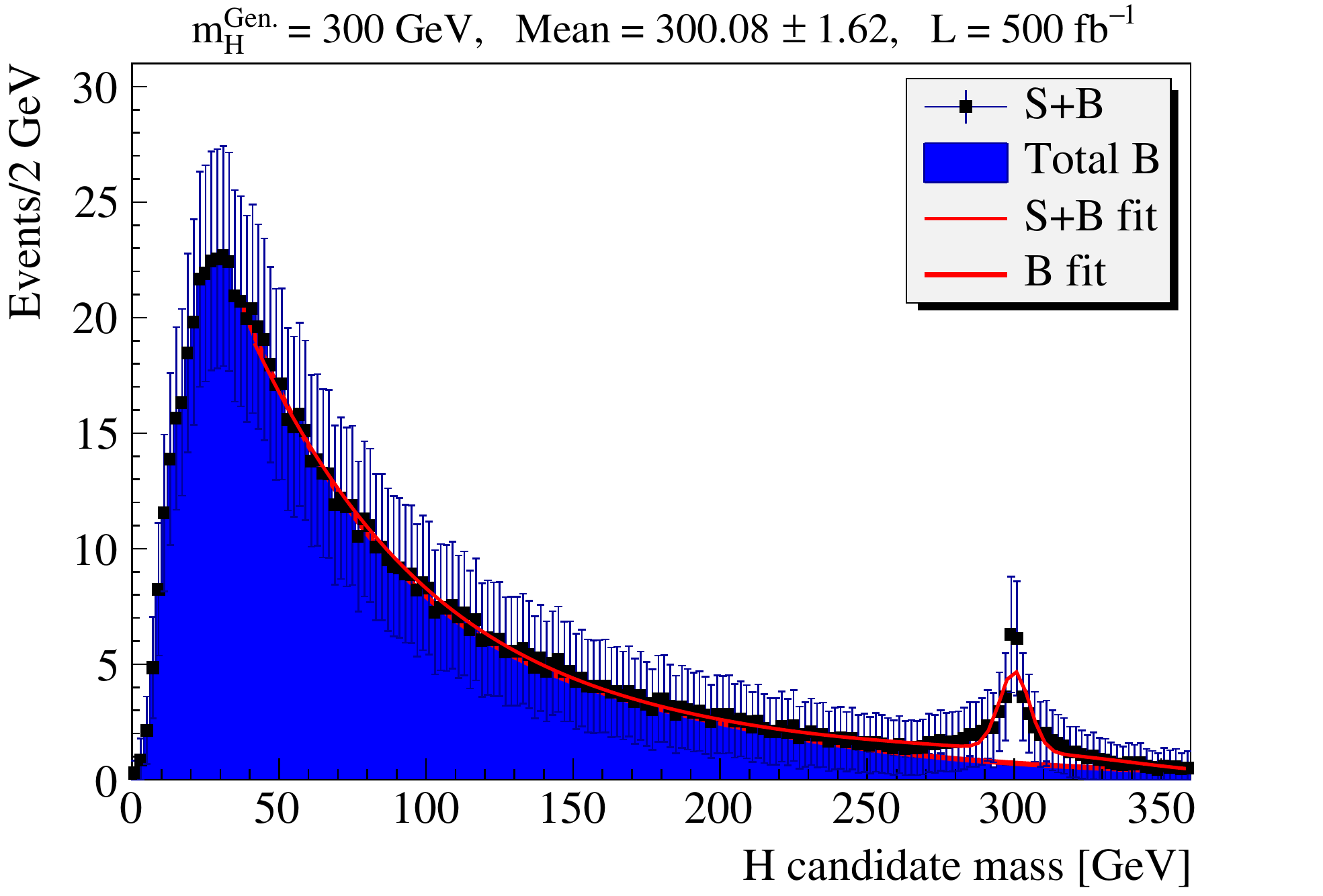}
    \caption{}
    \label{H300Poly.pdf}
    \end{subfigure} 
    \caption{Fitting results of the $H$ candidate mass distributions corresponding to the benchmark points a) BP3 and b) BP4. A polynomial function is fitted to the total background distribution. Statistical errors of the simulated data and the gaussian function ``mean'' parameter values are also shown.}
  \label{H2fit}
\end{figure}

\begin{figure}[h!]
  \centering
   \bigskip   
    \begin{subfigure}[b]{0.8\textwidth}
    \centering
    \includegraphics[width=\textwidth]{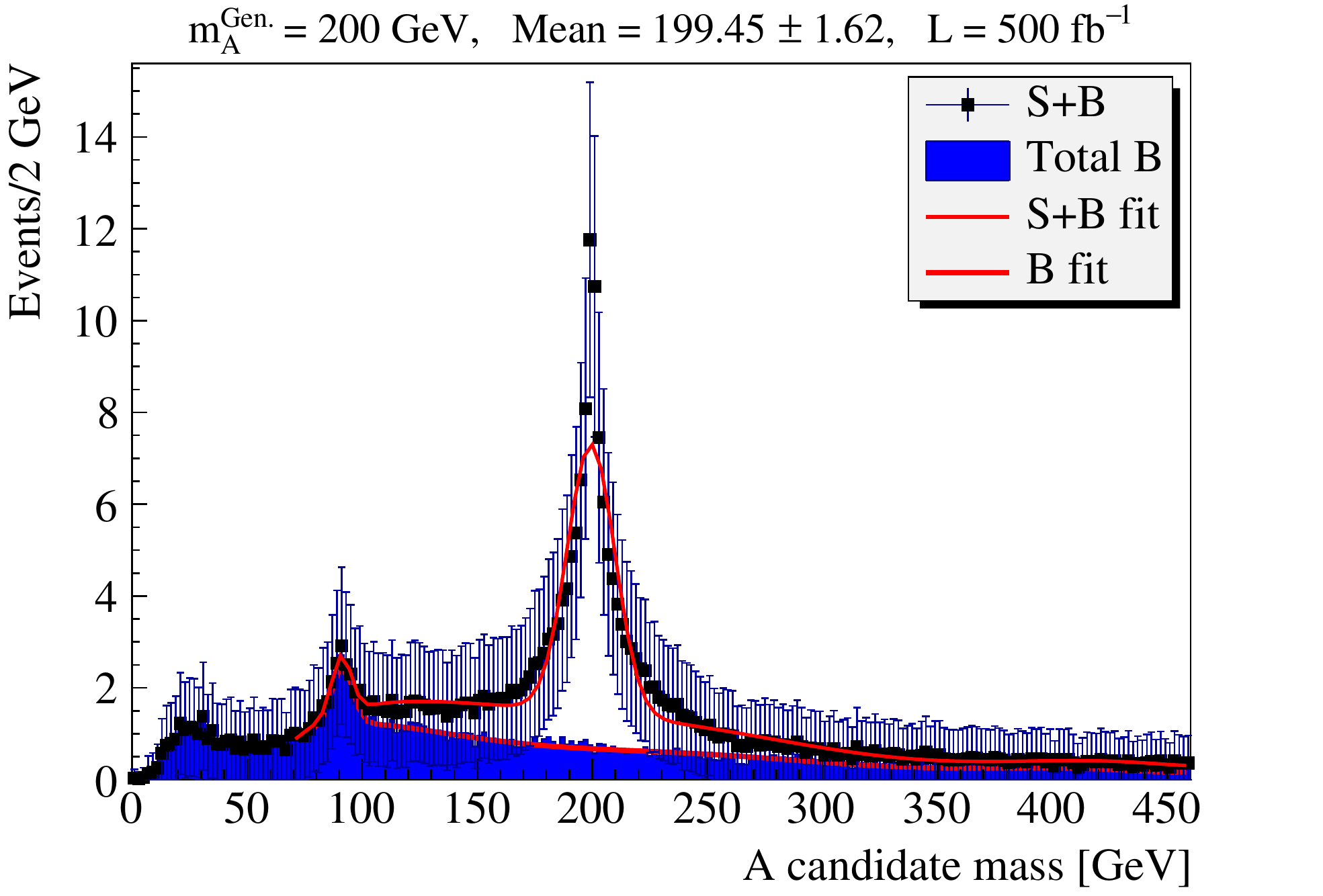}
    \caption{}
    \label{A200Poly.pdf}
    \end{subfigure}
    \bigskip
    \quad    
    \begin{subfigure}[b]{0.8\textwidth}
    \centering
    \includegraphics[width=\textwidth]{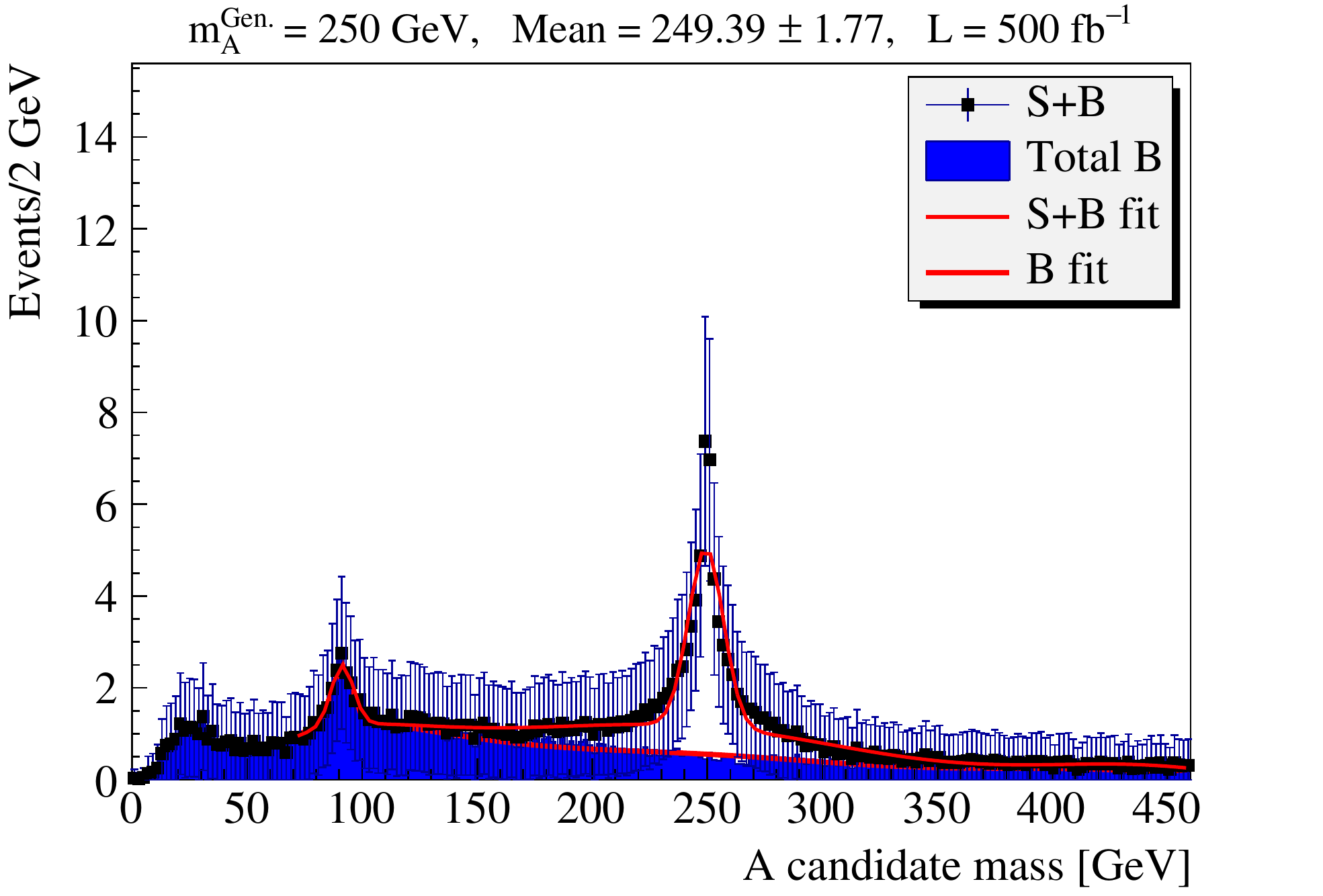}
    \caption{}
    \label{A250Poly.pdf} 
    \end{subfigure} 
\caption{Fitting results of the $A$ candidate mass distributions corresponding to the benchmark points a) BP1 and b) BP2. A polynomial function is fitted to the total background distribution. Statistical errors of the simulated data and the gaussian function ``mean'' parameter values are also shown.}
  \label{A1fit}
\end{figure}

\begin{figure}[h!]
  \bigskip
  \centering     
    \begin{subfigure}[b]{0.8\textwidth}
    \centering
    \includegraphics[width=\textwidth]{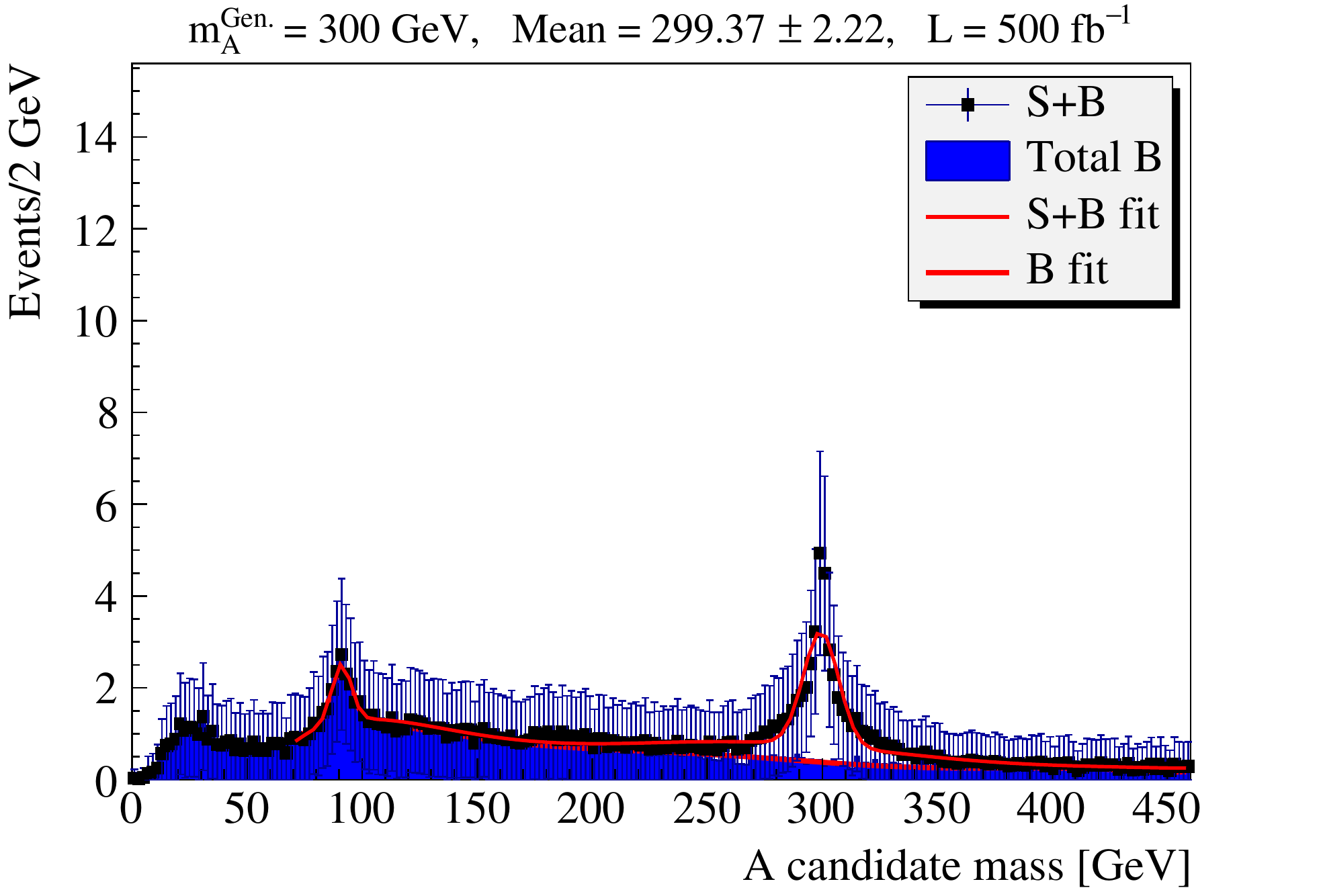}
    \caption{}
    \label{A300Poly.pdf}
    \end{subfigure} 
    \bigskip
    \quad    
    \begin{subfigure}[b]{0.8\textwidth}
    \centering
    \includegraphics[width=\textwidth]{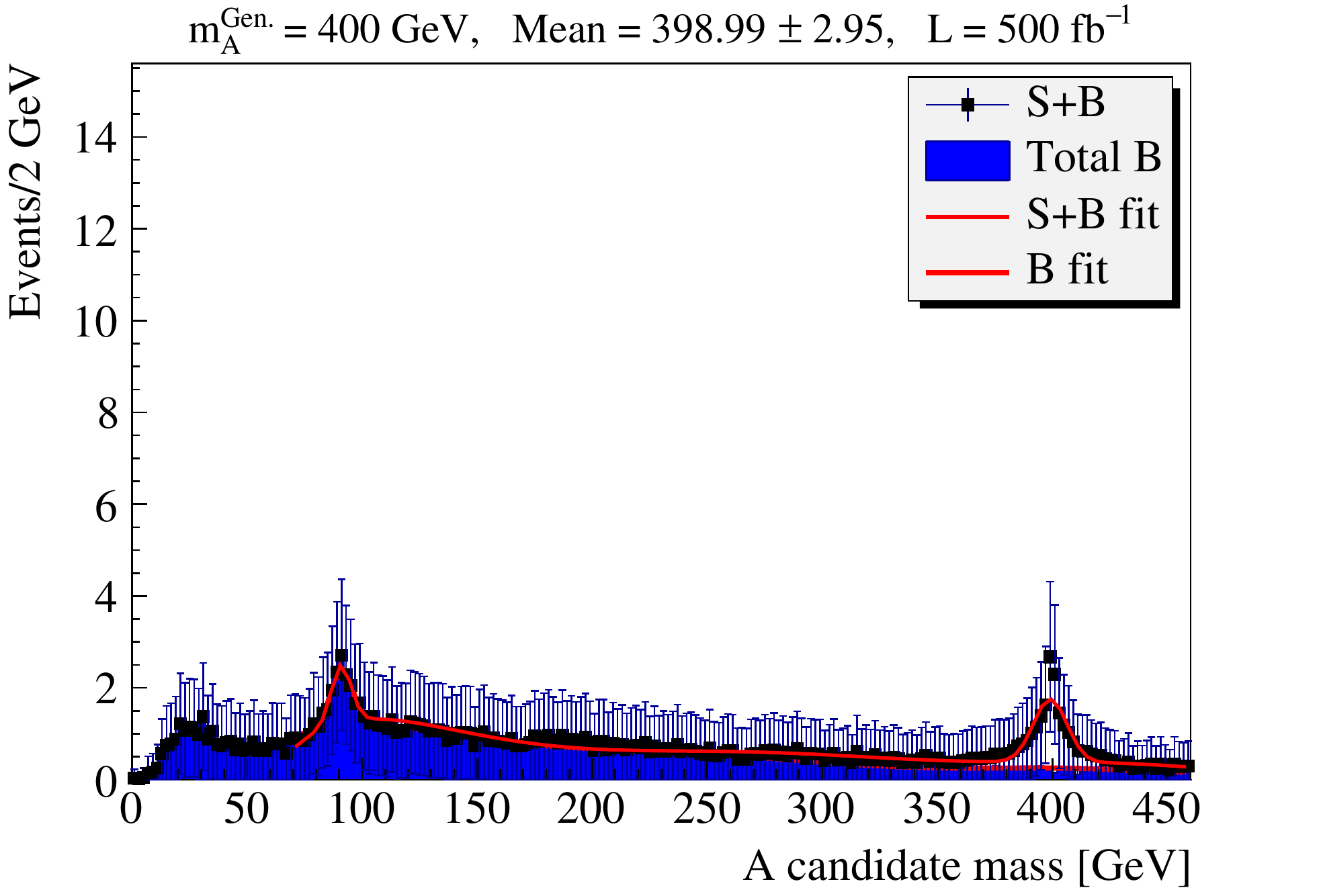}
    \caption{}
    \label{A400Poly.pdf}
    \end{subfigure} 
    \caption{Fitting results of the $A$ candidate mass distributions corresponding to the benchmark points a) BP3 and b) BP4. A polynomial function is fitted to the total background distribution. Statistical errors of the simulated data and the gaussian function ``mean'' parameter values are also shown.}
  \label{A2fit}
\end{figure}

Using the $H$ and $A$ candidate mass distributions of Fig. \ref{HA-allmHmasses}, reconstructed masses of the Higgs bosons are obtained as follows. Fitting an appropriate function to the mass distributions by ROOT 5.34 \cite{root}, reconstructed masses can be read from a certain fit parameter. The combination of a polynomial function and a gaussian function is used as the fit function for $H$ mass distributions. The gaussian part covers mainly the Higgs peak and thus the value of the ``mean'' parameter of the gaussian function provides the Higgs reconstructed mass. The fit function for $A$ mass distributions includes one more gaussian function to cover the small peak due to the $Z\gamma$ process (almost centered at the $Z$ boson mass). Figs. \ref{H1fit}-\ref{A2fit} show the fitting results.

\begin{figure}[h!]
  \centering
    \begin{subfigure}[b]{0.59\textwidth}
    \centering  
    \includegraphics[width=\textwidth]{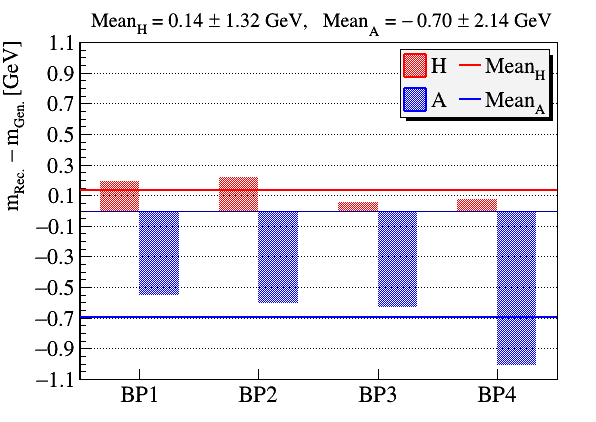}
    \caption{} 
    \label{recMinusGen} 
    \end{subfigure}
        \quad    
    \begin{subfigure}[b]{0.59\textwidth}
    \centering
    \includegraphics[width=\textwidth]{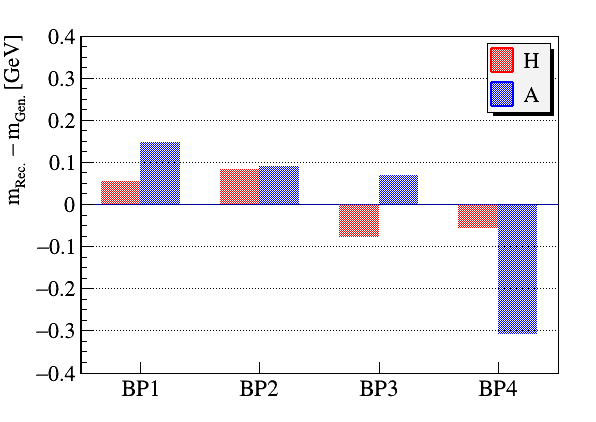}
    \caption{}
    \label{corr}
    \end{subfigure} 
  \caption{Differences between reconstructed and generated masses of the Higgs bosons $H$ and $A$ corresponding to different benchmark points a) before applying the off-set correction and b) after applying the off-set correction.}
  \label{recMinusGencorr}
\end{figure}

``Mean'' values of the gaussian part of the fit function are taken as the reconstructed masses of the Higgs bosons. Considering the mean values shown in Figs. \ref{H1fit}-\ref{A2fit}, a small difference can be seen between the reconstructed masses and generated masses of the Higgs bosons. Fig. \ref{recMinusGen} provides the differences corresponding to different benchmark points for both $H$ and $A$ Higgs bosons.
Reconstructed masses must be in principle equal to generated masses. However, errors in energy, momentum and flight directions of the particles, mis-identification of jets, errors arising out of the fitting method, etc., are error sources which give rise to errors in reconstructed masses. Optimization of the jet reconstruction algorithm and the fitting method may help reduce the errors. Optimization of the jet algorithm can be done by comparing the resultant reconstructed jets and the generated particles with the help of MC truth matching tools. In case of a real experiment, there are some other potential error sources like electronic noise, underlying-events, pile up, etc., which may degrade the results. Hence, a careful correction concerning all such error sources must also be performed. 

As mentioned earlier, the fit function used for $A$ mass distributions has one more gaussian function which covers the small peak due to the $Z\gamma$ process. Mean value of this gaussian function gives the mass corresponding to this peak. Fitting results show that the average mass corresponding to this peak is $91.06$ GeV which is close to the $Z$ boson mass as expected.

In this work, a simple off-set correction is applied to the obtained Higgs reconstructed masses to reduce the errors. To do so, a flat function is fitted to the plot of Fig. \ref{recMinusGen} to find the average difference between the reconstructed and generated masses for $H$ and $A$ Higgs bosons. As shown in the plot, average differences corresponding to the Higgs bosons $H$ and $A$ are $0.14$ and $-0.70$ respectively. To apply the off-set correction, $H$ reconstructed masses are decreased by 0.14 GeV and $A$ reconstructed masses are increased by 0.70 GeV. Corrected reconstructed masses are provided in table \ref{HAcormasstab}. 
\begin{table}[h]
\normalsize
\fontsize{11}{7.2} 
    \begin{center}
         \begin{tabular}{ >{\centering\arraybackslash}m{.18in} >{\centering\arraybackslash}m{1.2in}  >{\centering\arraybackslash}m{.85in}  >{\centering\arraybackslash}m{.85in} >{\centering\arraybackslash}m{.85in}  >{\centering\arraybackslash}m{.85in}  >{\centering\arraybackslash}m{.85in} } 
  &  & \cellcolor{blizzardblue}{BP1} & \cellcolor{blizzardblue}{BP2} & \cellcolor{blizzardblue}{BP3} & \cellcolor{blizzardblue}{BP4} \parbox{0pt}{\rule{0pt}{1ex+\baselineskip}}\\ \Xhline{3\arrayrulewidth}
  \multirow{2}{*}[-3.1pt]{\textbf{H}} &   \cellcolor{blizzardblue}{Gen. mass [GeV]} & 150 & 200 & 250 & 300 \parbox{0pt}{\rule{0pt}{1ex+\baselineskip}}\\ 
  &   \cellcolor{blizzardblue}{Rec. mass [GeV]} & 150.05$\pm$2.41 & 200.08$\pm$2.55 & 249.92$\pm$2.67 & 299.94$\pm$2.94 \parbox{0pt}{\rule{0pt}{1ex+\baselineskip}}\\  \Xhline{2\arrayrulewidth}

 \multirow{2}{*}[-3.1pt]{\textbf{A}} &   \cellcolor{blizzardblue}{Gen. mass [GeV]} & 200 & 250 & 300 & 400 \parbox{0pt}{\rule{0pt}{1ex+\baselineskip}}\\ 
  &   \cellcolor{blizzardblue}{Rec. mass [GeV]} & 200.15$\pm$3.76 & 250.09$\pm$3.91 & 300.07$\pm$4.36 & 399.69$\pm$5.09  \parbox{0pt}{\rule{0pt}{1ex+\baselineskip}}\\ \Xhline{3\arrayrulewidth}
        \end{tabular}
\caption{Generated and reconstructed masses of the Higgs bosons $H$ and $A$ with associated uncertainties. \label{HAcormasstab}}
  \end{center} 
\end{table} 
The difference between the reconstructed and generated masses after performing the off-set correction is also shown in Fig. \ref{corr}. Results of this Fig. show that differences corresponding to different benchmark points for $H$ ($A$) mass is smaller than $\sim 0.1$ ($\sim 0.3$) GeV. As indicated by the results of table \ref{HAcormasstab}, for all of the assumed benchmark points, $H$ and $A$ masses can be measured with few GeV uncertainty which is a statistical error. The uncertainty in a real experiment, however, is larger due to the systematic errors arising from various sources. Jet energy scale and resolution, particle momentum resolution, uncertainty arising from the fit function used to find the probability distribution function, etc., are main sources of uncertainty. 

\section{Signal significance} 
\begin{figure}[h!]
  \centering
  \includegraphics[width=0.59\textwidth]{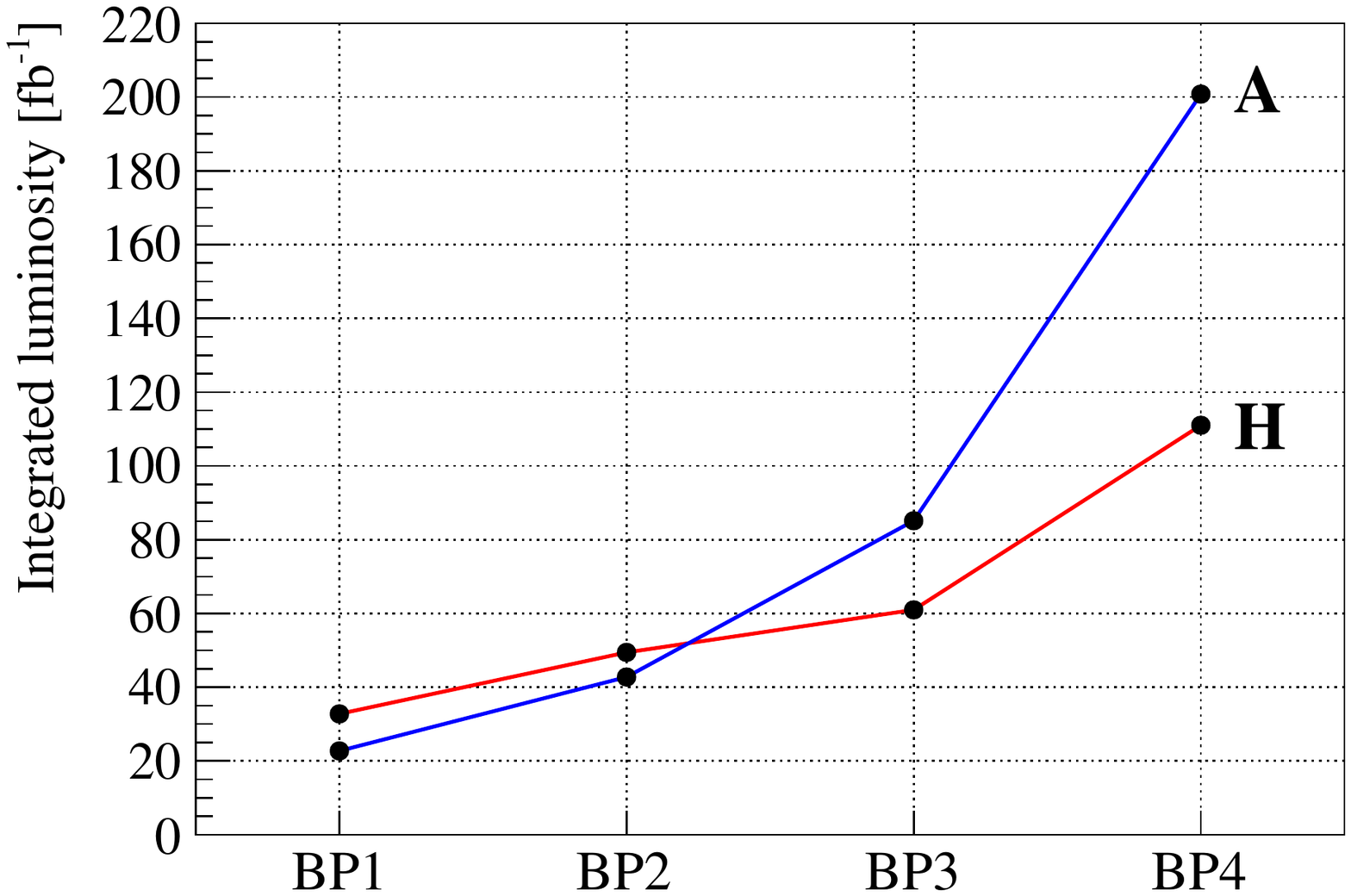}
  \caption{Required integrated luminosities for the Higgs bosons $H$ and $A$ to be observable with $5\sigma$ signals assuming different benchmark points.} \label{minL}
\end{figure}

To assess the observability of the Higgs bosons, signal significance corresponding to different mass distributions of Fig. \ref{HA-allmHmasses} are computed by first applying a mass window cut to distributions and then counting the number of signal and background Higgs candidate masses. Mass window cuts corresponding to different benchmark points are determined independently by optimizing the signal significance so that the signal significance has its maximum possible value for the chosen mass window cut. Computation is based on the integrated luminosity of $500$ $fb^{-1}$. Although the mass distributions and the signal significances are obtained at the integrated luminosity of $500$ $fb^{-1}$, both of the Higgs bosons are observable with $5\sigma$ signals at lower integrated luminosities. So, for each benchmark point, the integrated luminosity at which the Higgs boson is observable with a $5\sigma$ signal is computed and provided in table \ref{sigtab} ($5\sigma$ integrated L.) and also in Fig. \ref{minL}. Table \ref{sigtab} also provides mass window cuts and their associated efficiencies, signal total efficiencies, number of signal and background Higgs candidates and their ratio, and signal significances. 
\begin{table}[h!]
\normalsize
\fontsize{11}{7.2} 
    \begin{center} 
         \begin{tabular}{ >{\centering\arraybackslash}m{.2in} >{\centering\arraybackslash}m{1.50in}  >{\centering\arraybackslash}m{.6in}  >{\centering\arraybackslash}m{.6in} >{\centering\arraybackslash}m{.6in}  >{\centering\arraybackslash}m{.6in}  >{\centering\arraybackslash}m{.6in} } 
  &  & \cellcolor{blizzardblue}{BP1} & \cellcolor{blizzardblue}{BP2} & \cellcolor{blizzardblue}{BP3} & \cellcolor{blizzardblue}{BP4} \parbox{0pt}{\rule{0pt}{1ex+\baselineskip}}\\ \Xhline{3\arrayrulewidth} 
   &  \cellcolor{blizzardblue}{Gen. mass [GeV]} & 150 & 200 & 250 & 300 \parbox{0pt}{\rule{0pt}{1ex+\baselineskip}}\\ 
 \multirow{9}{*}[-7.7pt]{\textbf{H}}   & \cellcolor{blizzardblue}{Mass window [GeV]} & 136-175 & 189-219 & 241-265 & 292-310 \parbox{0pt}{\rule{0pt}{1ex+\baselineskip}}\\ 
  & \cellcolor{blizzardblue}{Mass window cut eff.} & 0.469 & 0.434 & 0.417 & 0.336 \parbox{0pt}{\rule{0pt}{1ex+\baselineskip}}\\ 
  &  \cellcolor{blizzardblue}{Total eff.} & 0.367 & 0.347 & 0.331 & 0.294 \parbox{0pt}{\rule{0pt}{1ex+\baselineskip}}\\ 
  &  \cellcolor{blizzardblue}{$S$} & 172.7 & 94.5 & 55.4 & 25.9  \parbox{0pt}{\rule{0pt}{1ex+\baselineskip}}\\ 
  &  \cellcolor{blizzardblue}{$B$} & 78.1 & 35.3 & 15.0 & 6.0  \parbox{0pt}{\rule{0pt}{1ex+\baselineskip}}\\  
&\cellcolor{blizzardblue}{$S/B$} & 2.2 & 2.7 & 3.7 & 4.3  \parbox{0pt}{\rule{0pt}{1ex+\baselineskip}}\\ 
&\cellcolor{blizzardblue}{$S/\sqrt{B}$} & 19.5 & 15.9 & 14.3 & 10.6  \parbox{0pt}{\rule{0pt}{1ex+\baselineskip}}\\ 
&  \cellcolor{blizzardblue}{Integrated L. [$fb^{-1}]$} &  \multicolumn{4}{c}{500} \parbox{0pt}{\rule{0pt}{1ex+\baselineskip}}\\ 
&  \cellcolor{blizzardblue}{$5\sigma$ integrated L. [$fb^{-1}]$} & 32.8 & 49.4 & 61.0 & 111.0  \parbox{0pt}{\rule{0pt}{1ex+\baselineskip}}\\ 
\Xhline{3\arrayrulewidth} 
 &  \cellcolor{blizzardblue}{Gen. mass [GeV]} & 200 & 250 & 300 & 400 \parbox{0pt}{\rule{0pt}{1ex+\baselineskip}}\\ 
  \multirow{9}{*}[-7.7pt]{\textbf{A}} & \cellcolor{blizzardblue}{Mass window [GeV]} & 187-217 & 239-267 & 292-311 & 386-414 \parbox{0pt}{\rule{0pt}{1ex+\baselineskip}}\\ 
  & \cellcolor{blizzardblue}{Mass window cut eff.} & 0.433 & 0.427 & 0.367 & 0.399 \parbox{0pt}{\rule{0pt}{1ex+\baselineskip}}\\ 
  &  \cellcolor{blizzardblue}{Total eff.} & 0.328& 0.322 & 0.272 & 0.324 \parbox{0pt}{\rule{0pt}{1ex+\baselineskip}}\\ 
  &  \cellcolor{blizzardblue}{$S$} & 77.1 & 43.9 & 22.8 & 14.3  \parbox{0pt}{\rule{0pt}{1ex+\baselineskip}}\\ 
  &  \cellcolor{blizzardblue}{$B$} & 10.8 & 6.6 & 3.5 & 3.3  \parbox{0pt}{\rule{0pt}{1ex+\baselineskip}}\\  
&\cellcolor{blizzardblue}{$S/B$} & 7.1 & 6.7 & 6.5 & 4.4  \parbox{0pt}{\rule{0pt}{1ex+\baselineskip}}\\ 
&\cellcolor{blizzardblue}{$S/\sqrt{B}$} & 23.5 & 17.1 & 12.1 & 7.9  \parbox{0pt}{\rule{0pt}{1ex+\baselineskip}}\\ 
&  \cellcolor{blizzardblue}{Integrated L. [$fb^{-1}]$} &  \multicolumn{4}{c}{500} \parbox{0pt}{\rule{0pt}{1ex+\baselineskip}}\\
&  \cellcolor{blizzardblue}{$5\sigma$ integrated L. [$fb^{-1}]$} & 22.7 & 42.8 & 85.1 & 200.8  \parbox{0pt}{\rule{0pt}{1ex+\baselineskip}}\\ 
\Xhline{3\arrayrulewidth}
        \end{tabular}
\caption{Optimized mass window cuts and corresponding efficiencies, signal total efficiencies, number of signal and background Higgs candidates after all selection cuts and mass window cut, signal to background ratios, signal significances, integrated luminosity at which the results are obtained, and the integrated luminosities at which the Higgs boson is observable with a $5\sigma$ signal ($5\sigma$ integrated L.).} 
 \label{sigtab}
  \end{center}
\end{table}
 
Results of table \ref{sigtab} show that, for all of the assumed benchmark points, the Higgs bosons $H$ and $A$ are observable with signals exceeding $5\sigma$ at integrated luminosities of $111$ and $201\, \, fb^{-1}$ respectively. Such luminosities are easily accessible to future linear colliders. According to the plot of Fig. \ref{minL}, as the Higgs bosons get heavier, the required integrated luminosities for obtaining $5\sigma$ signals increase. This is expected since according to table \ref{sXsec}, signal cross section decreases as the Higgs masses increase. Consequently, larger luminosity is needed to collect enough data.
 
\section{Conclusions} 
Working in the framework of the type-\RN{1} 2HDM (SM-like scenario), the question of observability of the heavy neutral CP-even and CP-odd Higgs bosons $H$ and $A$ at a linear collider operating at $\sqrt{s}=1$ TeV was addressed. The production process $e^- e^+ \rightarrow A H$ was assumed, where the produced pseudoscalar Higgs $A$ experiences the decay channel $A\rightarrow ZH$ followed by the leptonic ($e^-e^+$ or $\mu^-\mu^+$) decay of the $Z$ boson. Both of the resultant $H$ bosons are assumed to decay into a di-photon so that the signal can benefit from the enhancement due to the charged Higgs-mediated contribution to the $H$ di-photon decay at large $\tb$ values. Assuming four benchmark points in the mass parameter space of the 2HDM, signal and background events were generated, and taking advantage of the characteristics of the signal events, appropriate selection cuts were applied to events to enrich the signal. Momentum smearing was applied to leptons according to momentum resolution $\sigma_{p_T}/{p_T^2} = 2 \times 10^{-5}$ GeV$^{-1}$. Jet energy smearing and photon energy smearing are also performed according to energy resolutions $\sigma/E=3.5\, \%$ and $\sigma/E=2.7\, \%$ respectively. Mass distributions for both Higgs bosons $H$ and $A$ were obtained by the help of photon and lepton pairs invariant masses, and finally fitting a function to distributions, reconstructed masses of the Higgs bosons were obtained with few GeV uncertainty. Signal significances corresponding to different benchmark points were also computed by applying an optimized mass window cut. Results indicate that, for all of the assumed benchmark points, Higgs bosons $H$ and $A$ are observable with signals exceeding $5\sigma$ at integrated luminosities $111$ and $201\,\, fb^{-1}$ respectively. The required luminosities are easily accessible to future linear colliders. Mass measurement is also possible for all of the assumed benchmark points. The mass range in which the Higgs boson $H$ ($A$) is observable is 150-300 (200-400) GeV. 

\section*{Acknowledgements}
The analysis presented in this work was fully performed using the computing cluster at Shiraz University, college of sciences. We would like to thank Dr. Mogharrab for his careful maintenance and operation of the computing cluster. 

\bibliography{BIB_TO_USE}{}
\bibliographystyle{JHEP}  


\end{document}